\documentclass[5p,sort&compress]{elsarticle}
\usepackage{widetext}
\usepackage{multirow}
\usepackage{epstopdf}
\usepackage{subcaption,graphicx,float}
\usepackage{amsmath, amsfonts, amssymb}
\usepackage{amsthm}
\usepackage{yhmath}
\usepackage{epsf}
\usepackage{amsfonts}
\usepackage{stmaryrd}
\usepackage{amssymb}
\usepackage{leftidx}
\usepackage{xcolor}
\usepackage{mathtools}
\usepackage{placeins}
\usepackage{booktabs}
\usepackage{enumitem}
\usepackage{caption}
\usepackage{tabu}
\usepackage{multirow}
\usepackage{stackengine}
\usepackage[utf8]{inputenc}
\usepackage[english]{babel}
\usepackage{bm}
\usepackage{hyperref}
\hypersetup{
    colorlinks=true,
    linkcolor=blue,
    citecolor=blue,
    urlcolor=blue,
    linktoc=all
}
\usepackage{accents}
\usepackage{siunitx}

\DeclareMathAlphabet{\mathsfit}{T1}{\sfdefault}{\mddefault}{\sldefault}
\SetMathAlphabet{\mathsfit}{bold}{T1}{\sfdefault}{\bfdefault}{\sldefault}

\theoremstyle{plain}

\newdefinition{remark}{Remark}

\def\bfk{{\bf k}}

\def\bfu{{\bf u}}

\def\bfx{{\bf x}}
\def\bfy{{\bf y}}

\def\bfC{{\bf C}}

\def\bfI{{\bf I}}

\def\bfL{{\bf L}}

\def\bfS{{\bf S}}

\def\bfX{{\bf X}}
\def\bfF{{\bf F}}

\def\bfe{{\bf e}}

\def\Ktan{\mbox{\boldmath$\mathcal{K}$}}

\def\l{\lambda}

\def\e0{\varepsilon_0}

\def\oI{\overline{I}}

\def\obfF{\overline{\bfF}}

\def\obfS{\overline{\bfS}}
\def\ol{\overline{\lambda}}

\def\s0{\sigma_0}

\def\obfS{\overline{\bfS}}
\def\obfF{\overline{\bfF}}

\long\def\symbolfootnote[#1]#2{\begingroup%
\def\thefootnote{\fnsymbol{footnote}}\footnote[#1]{#2}\endgroup}

\begin{document}
\begin{frontmatter}

\title{The nonlinear elastic response of bicontinuous rubber blends\vspace{0.1cm}}

\vspace{-0.1cm}

\author[Illinois,Polytechnique]{Fabio Sozio}
\ead{fsozio@illinois.edu}

\author[Michelin]{Fran\c{c}ois Lallet}
\ead{francois.lallet@michelin.com}

\author[Michelin]{Antoine Perriot}
\ead{antoine.perriot@michelin.com}

\author[Illinois]{Oscar Lopez-Pamies}
\ead{pamies@illinois.edu}

\address[Illinois]{Department of Civil and Environmental Engineering, University of Illinois, Urbana--Champaign, IL 61801, USA }

\address[Polytechnique]{Solid Mechanics Laboratory, \'Ecole Polytechnique, 91128 Palaiseau, France}

\address[Michelin]{Manufacture Fran\c{c}aise de Pneumatiques Michelin, 63040 Clermont Ferrand, France \vspace{0.05cm}}

\begin{abstract}

\vspace{0.2cm}

Rubber blends are ubiquitous in countless technological applications. More often than not, rubber blends exhibit complex interpenetrating microstructures, which are thought to have a significant impact on their resulting macroscopic mechanical properties. As a first step to understand this potential impact, this paper presents a bottom-up or homogenization study of the nonlinear elastic response of the prominent class of bicontinuous rubber blends, that is, blends made of two immiscible constituents or phases segregated into an interpenetrating network of two separate but fully continuous domains that are perfectly bonded to one another. The focus is on blends that are isotropic and that contain an equal volume fraction (50/50) of each phase. The microstructures of these blends are idealized as microstructures generated by level cuts of Gaussian random fields that are suitably constrained to be periodic so as to allow for the construction of unit cells over which periodic homogenization can be carried out. The homogenized or macroscopic elastic response of such blends are determined both numerically via finite elements and analytically via a nonlinear comparison medium method. The numerical approach makes use of a novel meshing scheme that leads to conforming and periodic simplicial meshes starting from a voxelized representation of the microstructures. Results are presented for the fundamental case when both rubber phases are Neo-Hookean, as well as when they exhibit non-Gaussian elasticity. Remarkably, irrespective of the elastic behavior of the phases, the results show that the homogenized response of the blends is largely insensitive to the specific morphologies of the phases.

\vspace{0.2cm}

\endkeyword

\end{abstract}

\end{frontmatter}

\section{Introduction}\label{Sec:Intro}

Rubber blends have long been a staple in industries ranging from automotive to construction to consumer goods to healthcare. The reason behind their success is simple, rubber blending can result in new materials with significantly improved properties over those of the individual rubbers that are blended. As a prominent example, natural rubber is often combined with synthetic rubber in automotive tires so as to create a material with strong fracture resistance (due to the natural rubber) and a strong grip performance (due to the synthetic rubber). Depending on the extent of segregation of its constituents, rubber blends may range from perfectly \emph{compatible} --- wherein the constituents are mixed into one ``homogeneous'' phase --- to completely \emph{incompatible} --- wherein the constituents are arranged into microstructures. The microstructures of incompatible blends are often times comprised of complex interpenetrating networks, which are thought to have a significant impact on the resulting macroscopic mechanical and physical properties of the blends \cite{ryan2002designer,pernot2002design}.

\begin{figure*}[t!]
\centering
\includegraphics[width=.9\textwidth]{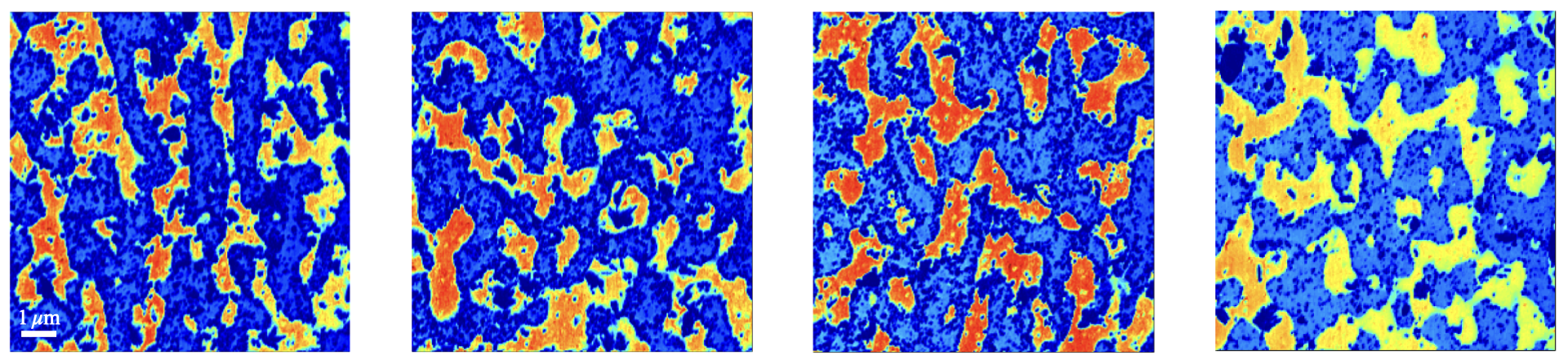}
\caption{AFM (atomic force microscopy) images of a typical isotropic 50/50 bicontinuous rubber blend used in automotive tires. Data courtesy of Michelin.}
\label{Fig1}
\end{figure*}

In this paper, as a first step to understand how the microstructures of rubber blends may affect their macroscopic mechanical properties, we carry out a bottom-up or homogenization study of the nonlinear elastic response of a prominent class of these materials, that of bicontinuous rubber blends. These are binary mixtures in which each constituent or phase is segregated into an interpenetrating network of two separate but fully continuous domains that are perfectly bonded to one another. For definiteness, we focus on blends that are isotropic and that contain an equal volume fraction (50/50) of each phase\footnote{In practice, 50/50 rubber blends are most often preferred so as to increase the probability of ending up with bicontinuous microstructures.}; see Fig. \ref{Fig1}.

By now, it is well established that the microstructures of rubber blends can be viewed as the result of spinodal decompositions and hence that they can be described, to within a reasonable degree of accuracy \cite{cabral2018spinodal, inguva2021continuum}, by the Cahn-Hilliard equation \cite{cahn1958free}. Accordingly, one can generate the microstructures of the blends of interest in this work by solving suitably calibrated Cahn-Hilliard equations; see, e.g., \cite{carolan2015co}. Alteratively, one can make use of the more computationally inexpensive strategy introduced by Cahn \cite{cahn1965phase}, which consists in generating approximately spinodal microstructures through level cuts of Gaussian random fields (GRF)\footnote{GRFs \cite{adler2010geometry} have long found extensive utility across a wide spectrum of fields and, more recently, have emerged as a powerful tool in the study of random heterogeneous media, such as microemulsions, porous media, metallic alloys, and polymer blends \cite{berk1987scattering, berk1991scattering, teubner1991level, roberts1995transport, torquato2002random, grigoriu2003random,soyarslan20183d,kumar2020inverse,Kochmann21,senhora2022optimally}.}. In this work, we make use of the latter approach.

The organization of the paper is as follows. We begin in Section \ref{Sec:Microstructures} by presenting the methodology utilized to construct the microstructures of bicontinuous rubber blends. In the footstep of a now well-established strategy \cite{Gusev97,MMS99,lopez2013nonlinear,GLLP23}, we approximate the isotropic microstructures of interest in this work as infinite media made out of the periodic repetition of a unit cell containing a random interpenetrating network of sufficient complexity that leads to approximately isotropic elastic behaviors. Such unit cells are first constructed (Subsections \ref{Sec:Generation} and \ref{Sec: Filter}) through level cuts of Gaussian random fields, constrained to be periodic, making use of a voxelized discretization. A novel scheme is then introduced (Subsection \ref{Sec:meshing})  that allows to convert the voxelized representation of the microstructures into conforming and periodic simplicial finite element (FE) discretizations. This conversion is crucial in order to describe the morphology of the phases faithfully\footnote{Most of the computational results available in the literature for the mechanical response of materials with bicontinuous microstructures are generated by making use either of \emph{non-conforming} voxels in fast-Fourier-transform (FFT) approaches or \emph{non-conforming} hexahedral meshes in FE approaches, which feature a large number of artificial corners at the interfaces between the two phases.} and to be able to solve the governing equations --- in our case, the equations of finite elastostatics --- accurately. In Section \ref{Sec:Mechanics}, we formulate the homogenization problem that defines the macroscopic nonlinear elastic response of the blends. In Sections \ref{Sec: Results 1}, \ref{Sec: Results 2}, and \ref{Sec: Results 3}, we then present numerical solutions for the homogenization problem in the limit of small deformations (when the response is linear elastic), for the basic case when the rubber phases are Neo-Hookean, and for the general case when the rubber phases feature non-Gaussian elasticity, respectively. Complementary to the numerical results presented in Sections \ref{Sec: Results 1} through \ref{Sec: Results 3}, we present in Section \ref{Sec: Approximation} an analytical approximate solution for the homogenization problem by means a nonlinear comparison medium technique \cite{lopez2013nonlinear}. We close in Section~\ref{Sec:FinalComments} by recording a number of final comments.

\section{Construction and discretization of the bicontinuous microstructures of rubber blends} \label{Sec:Microstructures}

Consider a rubber blend made of two constituents or phases, labeled $r=1$ and $2$, that in its initial (undeformed) configuration occupies the open domain $\Omega_0\in \mathbb{R}^3$, with boundary $\partial\Omega_0$. Denote by $\Omega_0^{(1)}$ and $\Omega_0^{(2)}$ the subdomains occupied individually by the two phases such that $\Omega_0=\Omega_0^{(1)}\cup\Omega_0^{(2)}$ and $\Omega_0^{(1)}\cap\Omega_0^{(2)}=\varnothing$. Identify material points in the blend by their initial position vector $\bfX\in\Omega_0$ and denote by $\theta_0^{(1)}(\bfX)$ and $\theta_0^{(2)}(\bfX)$ the characteristic or indicator functions describing the spatial locations occupied by each phase $r=1$ and $2$ in $\Omega_0$, that is,
\begin{equation*}
\theta_0^{(1)}(\bfX)=\left\{\begin{array}{ll}1 & \textrm{if }\bfX\in\Omega_0^{(1)}\\
0 &\textrm{else}\end{array}\right.
\end{equation*}
and
\begin{equation*}
\theta_0^{(2)}(\bfX)=\left\{\begin{array}{ll}1 & \textrm{if }\bfX\in\Omega_0^{(2)}\\
0 &\textrm{else}\end{array}\right. .
\end{equation*}
Following the strategy originally introduced by Cahn \cite{cahn1965phase}, consider in particular characteristic functions of the form
\begin{equation} \label{indicator-functions}
\theta_0^{(1)}(\bfX)= \mathcal{H}\left(f_0 - f(\bfX)\right)\;\, \textrm{and}\;\,
\theta_0^{(2)}(\bfX)= \mathcal{H}\left(f(\bfX) -f_0\right) ,
\end{equation}
where $\mathcal{H}(\cdot)$ stands for the Heaviside function, $f_0$ is a constant of choice, termed the level cut, and
\begin{equation} \label{grf-cahn-hilliard}
f(\mathbf{X}) = \frac{1}{\sqrt{2N}} \sum_{i=1}^N  \cos (\mathbf{k}_{i} \cdot \mathbf{X} + \phi_{i} ) .
\end{equation}
In this last expression, $\mathbf{k}_{i}$ stand for random wave vectors of the same magnitude $|\mathbf{k}_{i}|=k$ ($i=1,2,...,N$) and $\phi_{i}$ ($i=1,2,...,N$) are random phase angles in the range $[0,\pi]$.

\begin{figure*}[t!]
\centering
\includegraphics[width=0.9\textwidth]{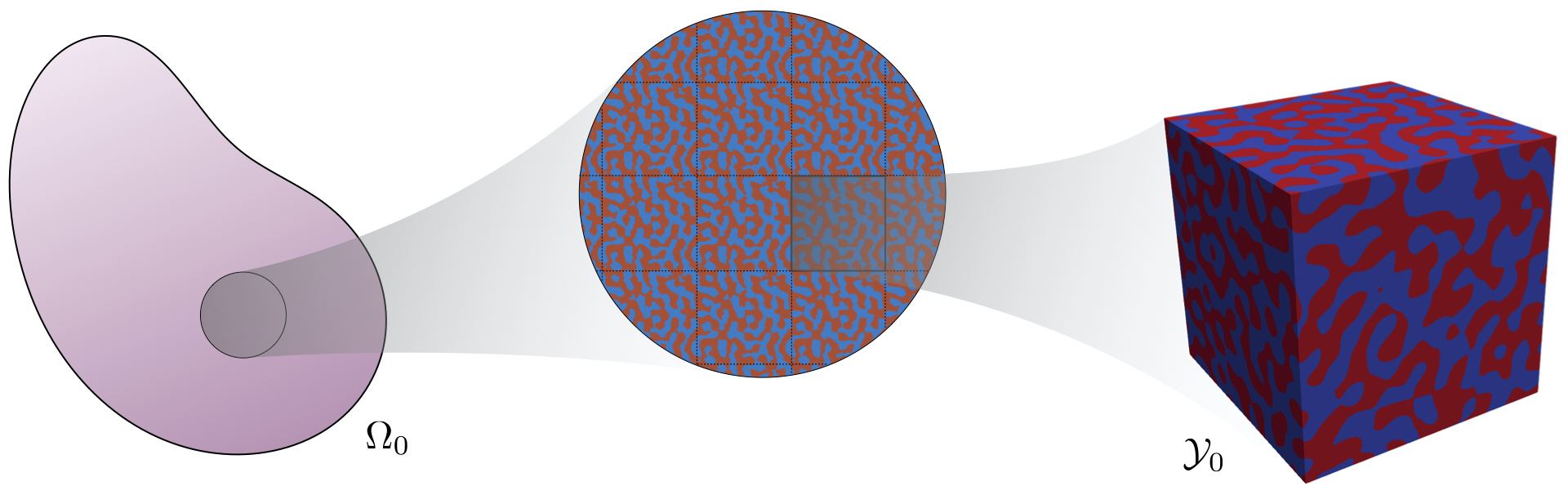}
\caption{Schematics (in the initial configuration) of a bicontinuous rubber blend and of the unit cell $\mathcal{Y}_0$ that defines its periodic microstructure.}
\label{Fig2}
\end{figure*}

\paragraph{The limit of infinitely many waves as $N\nearrow \infty$} In the limit as $N\nearrow \infty$, the function \eqref{grf-cahn-hilliard} converges to a statistically homogeneous, ergodic, and isotropic GRF; see, e.g., Chapter 8.2 in the monograph by Torquato \cite{torquato2002random}. In that limit, the value of the level cut $f_0$ in (\ref{indicator-functions}) determines the initial volume fractions
\begin{equation} \label{volume-fractions}
c_0^{(r)}:= \dfrac{1}{|\Omega_0|}\displaystyle\int_{\Omega_0}\theta_0^{(r)}(\bfX)\,{\rm d}\bfX\qquad r=1,2
\end{equation}
of the phases via the relations \cite{soyarslan20183d}
\begin{equation}  \label{volumefraction-levelcut}
f_0 = \sqrt{2} \operatorname{erf}^{-1}(2c_0^{(1)}-1) = \sqrt{2} \operatorname{erf}^{-1}(2c_0^{(2)}-1),
\end{equation}
where $\operatorname{erf}^{-1}(\cdot)$ stands for the inverse error function. Note that $f_0=0$ corresponds to a blend that contains an equal volume fraction $c_0^{(1)}=c_0^{(2)}=1/2$ of each phase, which is the case of interest here.

\paragraph{Periodicity} In order to be able to formulate a proper homogenization problem, we consider further that the characteristic functions (\ref{indicator-functions}) are periodic. In other words, we consider that the microstructure of the blend is the result of the periodic repetition of a unit cell, say $\mathcal{Y}_0$. For convenience, without loss of generality, we take $\mathcal{Y}_0=(0,1)^3$ with respect to the Cartesian laboratory frame of reference $\{\bfe_1,\bfe_2,\bfe_3\}$; see Fig. \ref{Fig2}. For this choice, it follows that the random wave vectors $\mathbf{k}_{i}$ in (\ref{grf-cahn-hilliard}) must be selected to be of the form \cite{soyarslan20183d}
\begin{equation} \label{periodic-components}
\mathbf{k}_i = 2\pi \,( m_i \bfe_1 + n_i \bfe_2 + p_i \bfe_3 )
\end{equation}
with
\begin{equation*}
\left\{\begin{array}{ll}m_i,n_i,p_i \in \mathbb{Z}\vspace{0.2cm}\\
m_i^2+n_i^2+p_i^2 = \left(\dfrac{k}{2\pi}\right)^2=H^2\end{array}\right.\quad i=1,2,...,N,
\end{equation*}
where $H^2$ is any positive integer in the set\footnote{This is nothing more than Legendre's three-square theorem; see, e.g., Part III in the monograph by Landau \cite{Landau21}.} $H^2\in\mathbb{N} \setminus \left\lbrace 4^a(7+8b), ~a,b\in\mathbb{N}\cup\{0\} \right\rbrace=\{1, 2, 3, 4, 5, 6, 8, 9, 10, 11, 12$, $13, 14, 16, 17,...\}$. Physically, the quantity $H^{-1}$ represents the wavelength of the waves in the random field (\ref{grf-cahn-hilliard}) and hence describes the characteristic length scale of the microstructure within the unit cell. Note that the periodic representation (\ref{periodic-components}) implies that the microstructure of the blend is no longer statistically isotropic. However, as demonstrated in Subsection \ref{Sec: Filter} below, it can be made approximately isotropic by using a sufficiently large value for the parameter $H$. Note also that the relations (\ref{volumefraction-levelcut}) do \emph{not} apply in this case, however, as demonstrated in Subsection \ref{Sec: Filter} below, they provide useful approximations so long as $H$ is sufficiently large.

\begin{remark} \label{Rem:modification}
There are a number of ways in which one can enrich the function (\ref{grf-cahn-hilliard}) to generate more complex spinodal-like microstructures. For instance, one can  consider a sum of waves with multiple wavenumbers (i.e., $N_1$ waves with wavenumber $k$, $N_2$ waves with wavenumber $2k$, and so on). In such an approach, we would rewrite
\begin{equation*} %\label{grf-cahn-hilliard-mod}
f(\mathbf{X}) =  \frac{1}{\sqrt{2N}}  \sum_{j=1}^{M} \sum_{i=1}^{N_j}  \cos ( j\, \mathbf{k}_{i} \cdot \mathbf{X} + \phi_{i} )
\end{equation*}
with $N = \sum_{j=1}^M N_j$. In this work, we restrict attention to the standard random field (\ref{grf-cahn-hilliard}).
\end{remark}

\subsection{Construction of the unit cells via a voxel discretization}\label{Sec:Generation}

Having determined the parametrization (\ref{indicator-functions}) with (\ref{grf-cahn-hilliard}) and (\ref{periodic-components}) for the characteristic functions $\theta_0^{(1)}(\bfX)$ and $\theta_0^{(2)}(\bfX)$ that describes the microstructure of the blend, we now present a methodology to construct the underlying unit cell $\mathcal{Y}_0$. To this end, note that the parametrization (\ref{indicator-functions}) with (\ref{grf-cahn-hilliard}) and (\ref{periodic-components}) contains $3+4N$ parameters, namely, $f_0$, $N$, $H$, $\phi_i$, $m_i$, $n_i$, and $p_i$, whose values need to be prescribed to be consistent with the requirements that the blend must: ($i$) contain equal volume fraction $c_0^{(1)}=c_0^{(2)}=1/2$ of each phase, ($ii$) satisfy the periodicity conditions (\ref{periodic-components}), and ($iii$) lead to approximately isotropic elastic behaviors.

In order to construct microstructures with equal volume fraction of each phase, we set the value of the level cut in (\ref{indicator-functions}) to
\begin{equation*}
f_0=0.
\end{equation*}
This leads to a blend with $c_0^{(1)}\approx 1/2$ and $c_0^{(2)}=1-c_0^{(1)}$. The precise volume fractions that are obtained with this choice can always be computed \emph{a posteriori}. We will come back to this important point in the next subsection.

\begin{figure*}[t!]
\centering
\includegraphics[width=.95\textwidth]{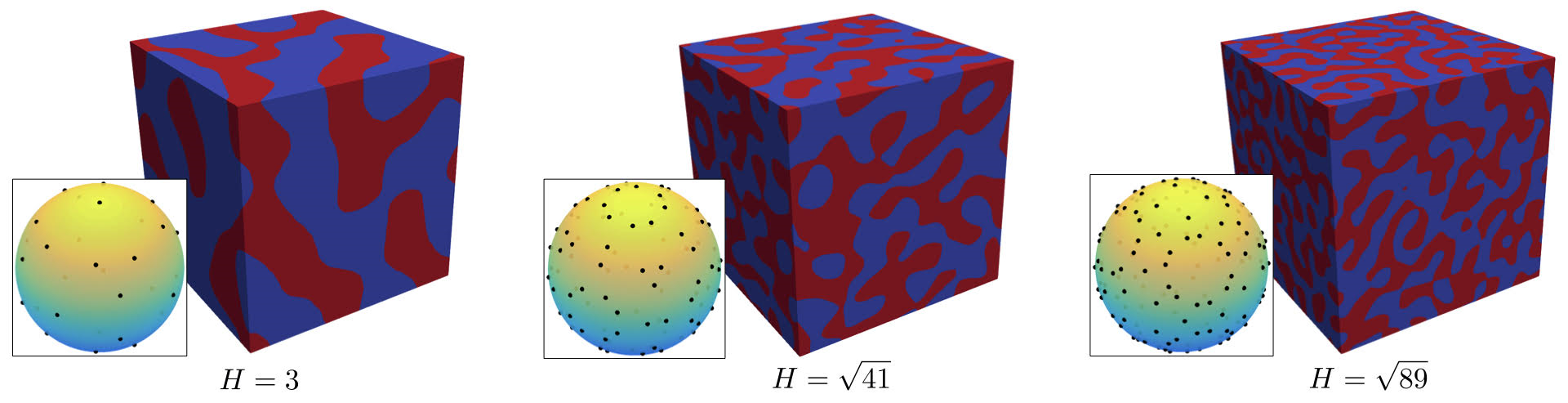}
\caption{Examples of unit cells $\mathcal{Y}_0$ for $f_0=0$, $N=1000$, and $H=3,\sqrt{41},\sqrt{89}$. The insets illustrate the corresponding admissible wave directions $(m_i,n_i,p_i)$ as points on the unit sphere.}
\label{Fig3}
\end{figure*}

As noted above, the function (\ref{grf-cahn-hilliard}) is only a GRF in the limit of infinitely many waves as $N\nearrow \infty$. In practice, $N$ can be set to a large but finite number. Numerical experiments suggest that for the microstructures of interest in this work it suffices to use
\begin{equation*}
N=1000.
\end{equation*}
This is the number of waves that we consistently use throughout this paper.

Given the periodicity constraint $m_i^2+n_i^2+p_i^2 = H^2$, it is clear that the value of the parameter $H$ constrains the range of triplets $(m_i,n_i,p_i)$ that are admissible. In particular, larger values of $H$ allow for a larger range of triplets $(m_i,n_i,p_i)$. Physically, the triplets $(m_i,n_i,p_i)$ describe the wave directions of the wave vectors $\bfk_i$. Therefore, a larger range of triplets $(m_i,n_i,p_i)$ implies a better approximation of an isotropic microstructure. By the same token, larger values of $H$ lead to better approximations of the isotropy of the blend. In this work, we consider the following three values
\begin{equation}
H=\left\{\begin{array}{l}\sqrt{9}=3 \vspace{0.2cm}\\ \sqrt{41} \approx 6.40 \vspace{0.2cm} \\ \sqrt{89} \approx 9.43\end{array}\right. .\label{H-values}
\end{equation}
A standard calculation shows that these values lead, respectively, to $30$, $96$, and $144$ different admissible triplets $(m_i,n_i,p_i)$.

Finally, we prescribe the values of the $4N$ random parameters $\phi_i$, $m_i$, $n_i$, and $p_i$ ($i=1,...,N$), subject to the constraints $\phi_i\in[0,\pi]$ and \eqref{periodic-components}, by making use of a random number generator. In our calculations, we make use of the \texttt{twister} algorithm with \texttt{shuffle} initializer in Matlab.

Once all the $3+4N$ parameters have been prescribed so that the random field $f(\bfX)$ can be evaluated at any material point $\bfX$ within the unit cell $\mathcal{Y}_0=(0,1)^3$, the next step is to generate a discretized description of the characteristic functions $\theta_0^{(1)}(\bfX)$ and $\theta_0^{(2)}(\bfX)$. We do so by discretizing the unit cell with a uniform array of $\lfloor 50 H\rfloor^3$ voxels, where $\lfloor \cdot\rfloor$ stands for the floor function. For the three values (\ref{H-values}) of $H$, these correspond to discretizations of $150^3$, $320^3$, and $471^3$ voxels, respectively. If $\theta_0^{(1)}(\bfX_c)=1$ at the centroid $\bfX_c$ of a voxel, then we set $\theta_0^{(1)}(\bfX)=1$ for all material points $\bfX$ in that voxel; similarly for  $\theta_0^{(2)}(\bfX)$. Figure \ref{Fig3} provides examples of three unit cells generated in this manner for $H=3,\sqrt{41},\sqrt{89}$. The insets show pictorial representations of the corresponding $30$, $96$, $144$ admissible wave directions.

\subsection{Geometric filtering of the unit cells} \label{Sec: Filter}

The unit cells $\mathcal{Y}_0$ generated by the above-described scheme correspond to blends that are isotropic and contain an equal volume fraction of each phase \emph{only} approximately. In this subsection, we describe a filtering process that discards unit cells wherein the required 50/50 volume fraction and/or the isotropy are not sufficiently well approximated.

\paragraph{The geometric filter for the volume fractions}  Once a voxelized discretization of a unit cell $\mathcal{Y}_0$ has been generated, it is straightforward to compute the volume fraction of each of its phases via the specialization $c_0^{(r)}= \int_{\mathcal{Y}_0}\theta_0^{(r)}(\bfX)\,{\rm d}\bfX$ $(r=1,2)$ of the formulas (\ref{volume-fractions}) to periodic microstructures. Unit cells for which such a computation yields
\begin{equation*} %\label{volume-fractions}
c_0^{(1)}\notin[0.49,0.51]
\end{equation*}
are ruled out as inadmissible.

\begin{figure*}[t!]
\centering
\includegraphics[width=0.9\textwidth]{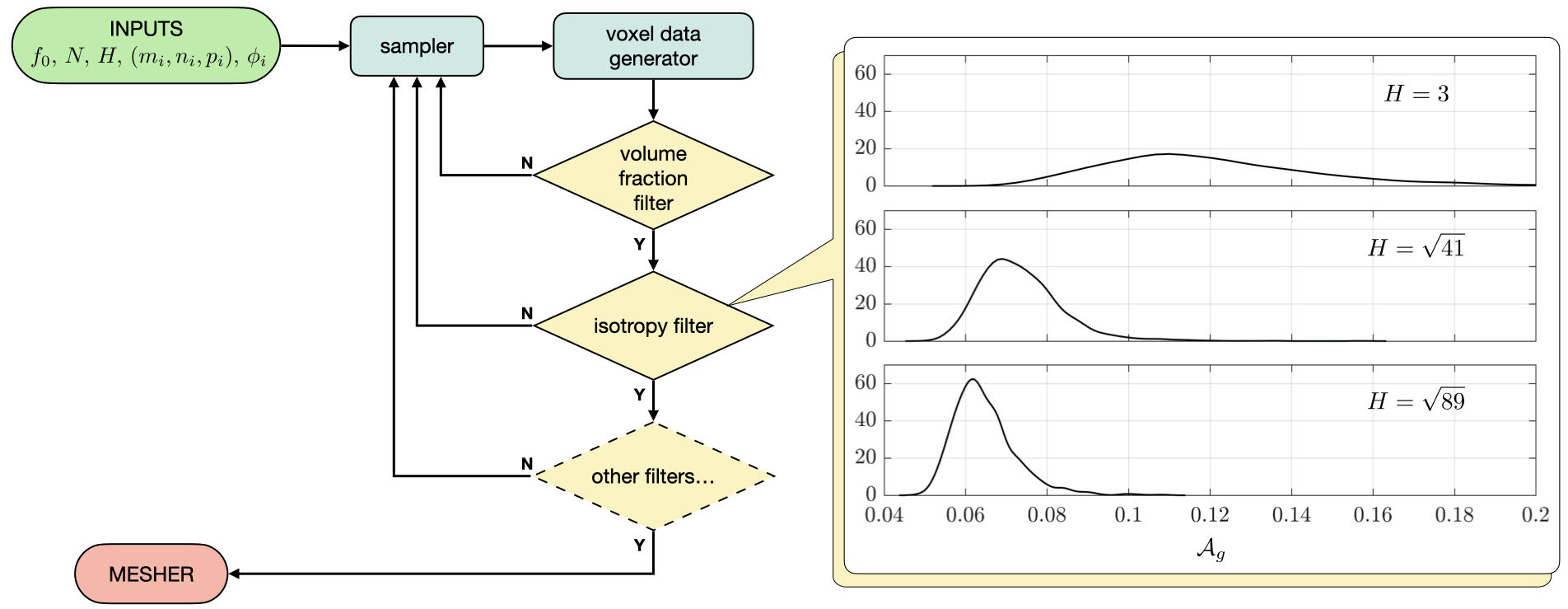}
\caption{Left: Flow chart of the algorithm used for the generation of the unit cells. Right: Frequencies of the unit cells that are generated by the algorithm as functions of their anisotropic measure $\mathcal{A}_g$. The results are shown for each of the three values (\ref{H-values}) of the parameter $H$.}
\label{Fig4}
\end{figure*}

\paragraph{The geometric filter for isotropy}  Once a voxelized discretization of a unit cell $\mathcal{Y}_0$ has been generated, it is also straightforward to compute the two-point correlation functions
\begin{equation} \label{two-point}
P^{(r)}_2(\bfX)=\displaystyle\int_{\mathcal{Y}_0}\theta_0^{(r)}(\bfX^\prime)\theta_0^{(r)}(\bfX+\bfX^\prime)\,{\rm d}\bfX^\prime
\end{equation}
$(r=1,2)$. Following in the footstep of \cite{lefevre2022curious}, as a first assessment of deviation from exact geometric isotropy, we compute the deviation of $P^{(r)}_2$ from its isotropic projection
\begin{align*} %\label{volume-fractions}
I^{(r)}_2(|\bfX|) = &\dfrac{1}{4\pi} \displaystyle\int_{0}^{\pi}\displaystyle\int_{0}^{2\pi}P^{(r)}_2\left(|\bfX| \cos\Theta \sin\Phi \,\bfe_1 +\right. \\
&\left. |\bfX|\sin\Theta \sin\Phi\,\bfe_2+|\bfX|\cos\Phi\,\bfe_3\right)\sin\Phi\, {\rm d}\Theta\,{\rm d}\Phi
\end{align*}
$(r=1,2)$ onto the space of functions that depend on $\bfX$ only through its magnitude $|\bfX|$. Unit cells that do not satisfy the condition
\begin{equation} \label{isotropic-filter}
\mathcal{A}_g=\dfrac{||P^{(1)}_2(\bfX)-I^{(1)}_2(\bfX)||_1}{c_0^{(1)}-(c_0^{(2)})^2}\leq 0.05,
\end{equation}
where $||\cdot||_1$ stands for the $L^1$ norm, are discarded as not sufficiently isotropic. Note that $\mathcal{A}_g$ is nothing more than a measure of the geometric anisotropy of the microstructure. Microstructures with $\mathcal{A}_g\leq 0.05$ are selected here as possibly sufficiently isotropic.

\begin{remark}
The integrals (\ref{two-point}) can be conveniently computed in Fourier space. We do so, in particular, by making use of the FFT algorithm built in Matlab.
\end{remark}
\begin{remark}
Whether the microstructures that satisfy the geometric isotropic filter (\ref{isotropic-filter}) lead to elastic behaviors that are sufficiently isotropic can only be determined by computing their elastic response. As demonstrated by the results presented in Sections \ref{Sec: Results 1}-\ref{Sec: Results 3} below, numerous microstructures that satisfy (\ref{isotropic-filter}) do indeed exhibit roughly isotropic elastic behaviors.
\end{remark}

Figure \ref{Fig4} summarizes in a flow chart the above-described algorithm for the construction and filtering of the unit cells. The figure also includes plots of the frequencies of the unit cells that are generated by the algorithm as functions of the anisotropic measure $\mathcal{A}_g$ for each of the three values (\ref{H-values}) of the parameter $H$. As anticipated above, larger values of $H$ clearly lead to microstructures that are more isotropic.

\subsection{Finite element discretization of the unit cells}\label{Sec:meshing}

While alluring for its simplicity, the construction of unit cells based on a voxel discretization --- as described in the preceding section --- has a major shortcoming: it leads to a description of the interfaces between the two rubber phases that contains a myriad of artificial corners. Depending on the physical problem of interest, the presence of such an artificial roughness may prevent the ability to solve the problem accurately. In this subsection, we present a scheme that allows to convert such voxel discretizations into conforming and periodic simplicial meshes that can then be used to solve the physical problem at hand --- in our case, again, finite elastostatics --- accurately via FEs\footnote{In a recent contribution, Hestroffer and Beyerlein \cite{Hestroffer2022} have introduced a scheme capable of converting voxel discretizations to conforming simplicial discretizations in the context of polycrystals. In its present form, however, the scheme does not allow to enforce periodicity.}.

\begin{figure*}[t!]
\centering
\includegraphics[width=0.9\textwidth]{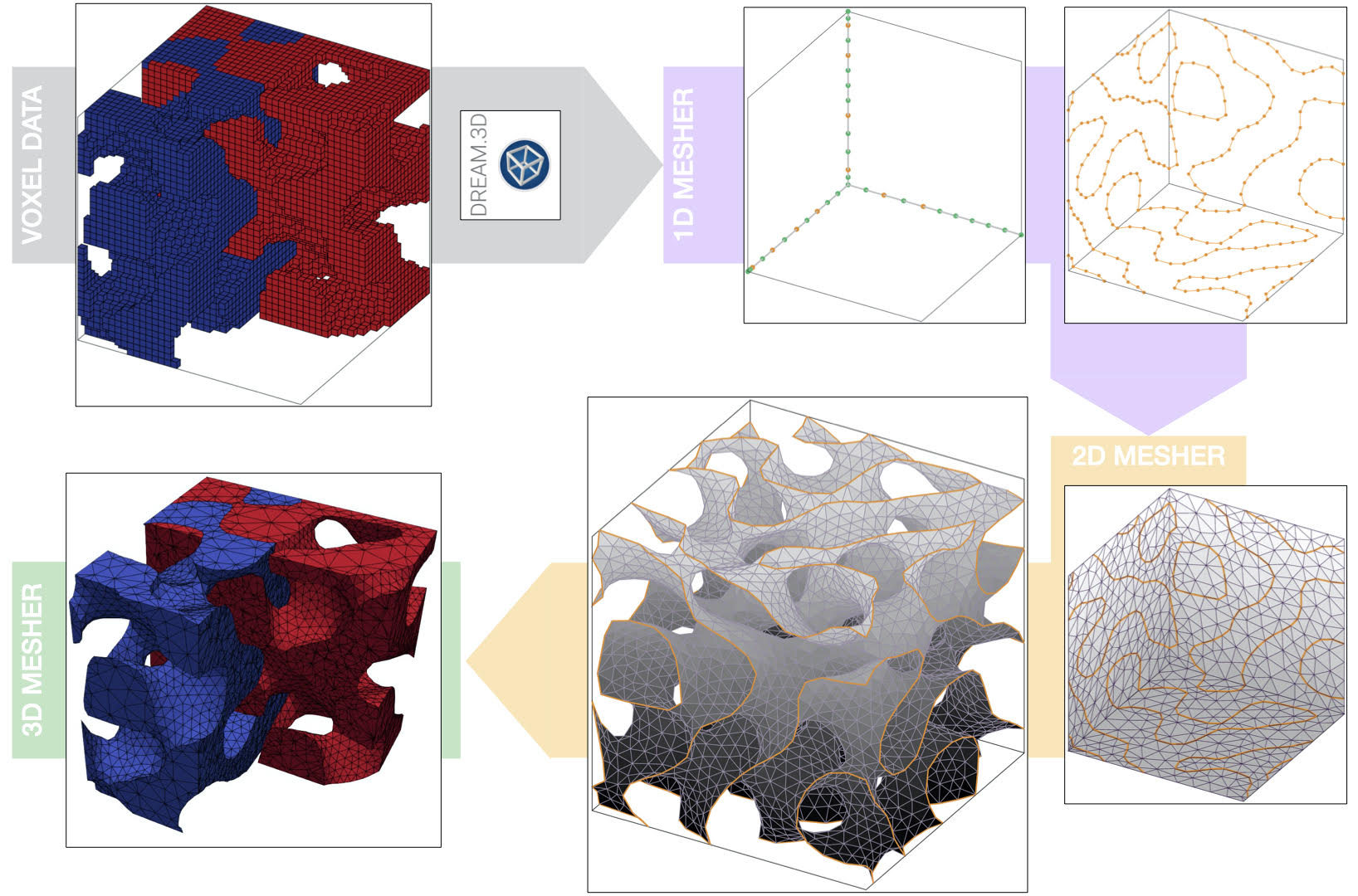}
\caption{Flow chart of the algorithm used to convert voxel discretizations of unit cells $\mathcal{Y}_0$ for bicontinuous rubber blends to conforming and periodic simplicial meshes.}
\label{Fig5}
\end{figure*}

As illustrated by Fig. \ref{Fig5}, the scheme is hierarchical and goes as follows. We begin by discretizing with 1D simplicial elements the 12 edges of the unit cell and the intersects of the interfaces between the rubber phases with the 6 facets of the unit cell. To enforce periodicity, this first step makes use of a master-slave approach. The next step consists in discretizing with 2D simplicial elements the 6 facets of the unit cell and the interfaces between the rubber phases within the unit cell in a manner that is conforming with the 1D simplicial elements generated in the first step. To so, we leverage tools from the computational geometry library CGAL \cite{cgal:eb-22b}, which support surface remeshing with fixed boundary edges and, consequently, allow to enforce periodicity. In particular, we make use of the \texttt{isotropic\_remeshing} function from the \texttt{Polygon\_mesh\_processing} package within CGAL. The conversion of the voxelized interfaces between the two rubber phases into 2D simplicial meshes is the most challenging step of the scheme. In a third step, we discretize the entirety of the unit cell with 3D simplicial elements starting from the 2D simplicial discretization generated in the second step.

The final step consists in assigning the appropriate rubber phase $r=1$ or 2 to each 3D simplicial element, as described by the characteristic functions $\theta_0^{(1)}(\bfX)$ and $\theta_0^{(2)}(\bfX)$. The procedure goes as follows. For a buffer $b=5h$, where $h$ stands for the average element diameter, an element is assigned to the rubber phase $r=1$ if $f(\bfX_c) < f_0 - b$ at the centroid $\bfX_c$ of that element. Similarly, an element is assigned to the rubber phase $r=2$ if $f(\bfX_c) > f_0 + b$. If $f_0 - b < f(\bfX_c) < f_0 + b$, the element is not initially assigned to any rubber phase. Unassigned elements that share a facet with an assigned element are assigned either to the same rubber phase as that shared element, if the shared facet does not belong to the interface between the rubbers, or to the other rubber phase if the shared facet belongs to the interface. This procedure is repeated iteratively until all elements are assigned to a rubber phase.\footnote{The simplicial discretizations that result from this procedure contain volume fractions $c_0^{(1)}$ and $c_0^{(2)}$ of the rubber phases that differ from those of the corresponding voxelized discretizations by a negligible amount.}

\begin{figure*}[t!]
\centering
\includegraphics[width=0.95\textwidth]{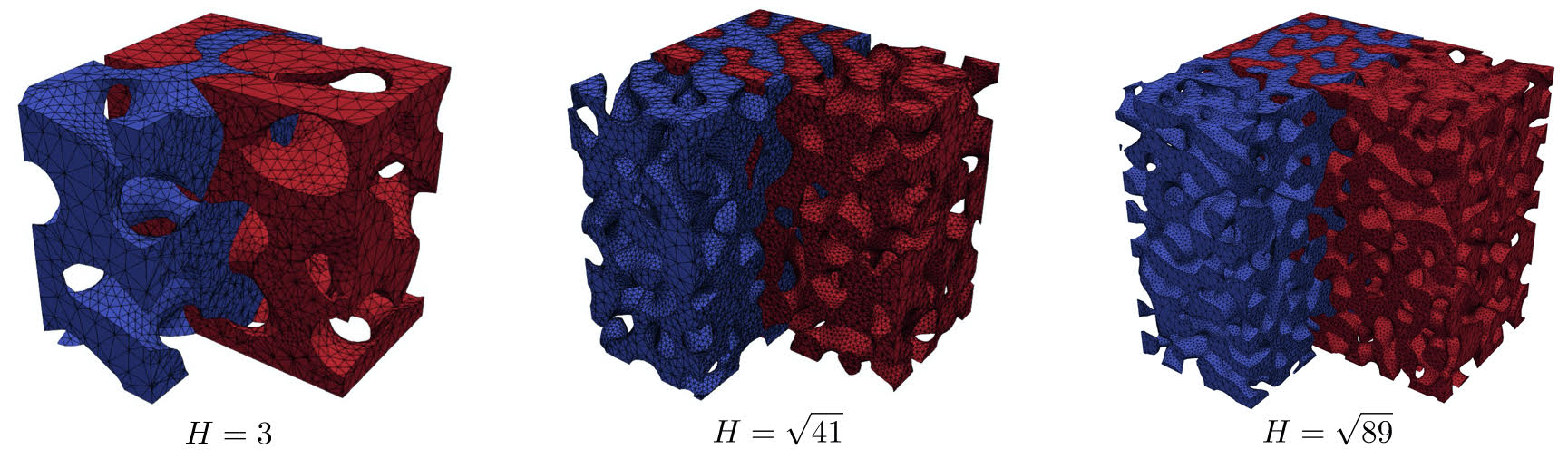}
\caption{Representative FE meshes (in the initial configuration) of unit cells $\mathcal{Y}_0$ of 50/50 bicontinuous rubber blends for the three values (\ref{H-values}) of the parameter $H$. The meshes are clipped in order to better illustrate the bicontinuous character of the microstructure.}
\label{Fig6}
\end{figure*}

Figure \ref{Fig6} presents representative FE meshes of unit cells for the three values (\ref{H-values}) of the parameter $H$. They contain about 60,000, 600,000, and 1.2 million elements, respectively.

\section{Formulation of the homogenization problem in finite elastostatics}\label{Sec:Mechanics}

Having defined the microstructure of the rubber blend, we are now in a position to formulate the homogenization problem that describes its macroscopic nonlinear elastic response when it is subjected to quasi-static finite deformations.

\subsection{The local problem}

\paragraph{Kinematics} In response to the applied boundary conditions described below, the position vector $\bfX$ of a material point in the blend will move to a new position specified by
\begin{equation*}
\bfx=\bfy(\bfX),
\end{equation*}
where $\bfy$ is a mapping from $\Omega_0$ to the current configuration $\Omega$. We consider only invertible deformations, and write the deformation gradient field at $\bfX$ as
\begin{equation*}
\bfF(\bfX)=\nabla\bfy(\bfX)=\frac{\partial\bfy}{\partial\bfX}(\bfX).
\end{equation*}

\paragraph{Constitutive behavior of the rubber phases} The two rubber phases in the blend are assumed to be isotropic and incompressible nonlinear elastic solids. In particular, we consider that their nonlinear elastic responses are characterized by stored-energy functions of the form
\begin{equation} \label{I1-based}
W^{(r)}(\bfF) = \left\{\begin{array}{ll}
\psi^{(r)}(I_1) & \textrm{if } J=1\vspace{0.2cm}\\
+\infty & \textrm{else}
\end{array}\right.
\end{equation}
$(r=1,2)$, where $I_1$ and $J$ stand for the invariants
\begin{equation*}
I_1=\textrm{tr}\,\bfC=\bfF\cdot\bfF=\lambda_1^2+\lambda_2^2+\lambda_3^2
\end{equation*}
and
\begin{equation*}
J=\sqrt{\det\bfC}=\det\bfF=\lambda_1\lambda_2\lambda_3
\end{equation*}
of the right Cauchy-Green deformation tensor $\bfC=\bfF^T\bfF$ and $\psi^{(r)}(I_1)$ are any functions of choice that satisfy the linearization conditions
\begin{equation}\label{linarization-cond}
\psi^{(r)}(3)=0,\qquad\dfrac{\textrm{d}\psi^{(r)}}{\textrm{d}I_1}(3)=\dfrac{\mu^{(r)}}{2},
\end{equation}
and the ellipticity conditions
\begin{align}\label{strong-ell-cond}
&\dfrac{\textrm{d}\psi^{(r)}}{\textrm{d}I_1}(I_1)>0,\nonumber\\
&\dfrac{\textrm{d}\psi^{(r)}}{\textrm{d}I_1}(I_1)+ 2 \left( I_1 - \lambda_k^2 - 2 \lambda_k^{-1}\right) \dfrac{\textrm{d}^2\psi^{(r)}}{\textrm{d}I_1^2}(I_1)> 0 \quad k=1,2,3
\end{align}
for all $I_1 \geq 3$ $(r=1,2)$. In these expressions, $\lambda_1$, $\lambda_2$, $\lambda_3$ stand for the singular values of the deformation gradient $\bfF$, while $\mu^{(r)}$ denote the initial shear moduli of the two rubber phases $r=1$ and $2$.

\begin{remark}
Stored-energy functions of the $I_1$-based form (\ref{I1-based}) are generalizations of the classical Neo-Hookean stored-energy function
\begin{equation}\label{W-Neo}
\psi^{(r)}(I_1)=\dfrac{\mu^{(r)}}{2}\left[ I_1-3 \right]
\end{equation}
that have been shown to describe reasonably well the response of a wide variety of unfilled and filled elastomers over large ranges of deformations, thus their use here; see, e.g., \cite{arruda1993three,beatty2003average,Benam21,lopez2010new,GSPLP15,MTOLP19,LWLPN2020}.
\end{remark}

\paragraph{Pointwise constitutive behavior of the blend} Granted the characteristic functions (\ref{indicator-functions}) and the stored-energy functions (\ref{I1-based}) for the two rubber phases, it follows that the first Piola-Kirchhoff stress tensor $\bfS$ at any material point $\bfX \in\Omega_0$ is given by the relation
\begin{align*} %\label{I1-stress-blend}
\bfS(\bfX)&=\dfrac{\partial W}{\partial\bfF}(\bfX,\bfF)-p\bfF^{-T}\nonumber\\
&=2\left[\theta_0^{(1)}(\bfX)\dfrac{\textrm{d}\psi^{(1)}}{\textrm{d}I_1}(I_1)+\theta_0^{(2)}(\bfX)\dfrac{\textrm{d}\psi^{(2)}}{\textrm{d}I_1}(I_1)\right]\bfF-p\bfF^{-T},
\end{align*}
where
\begin{equation}\label{W-local}
W(\bfX,\bfF)=\theta_0^{(1)}(\bfX)\psi^{(1)}(I_1)+\theta_0^{(2)}(\bfX)\psi^{(2)}(I_1)
\end{equation}
and $p$ stands for the arbitrary hydrostatic pressure associated with the incompressibility constraint $J=1$.

\paragraph{Governing equations} Neglecting inertia and body forces and, for simplicity of presentation, restricting attention to the Dirichlet boundary condition $\bfy(\bfX)=\overline{\bfy}(\bfX)$ over the entirety of the boundary $\partial\Omega_0$ of the rubber blend, the combination of all the above ingredients with the balance of linear momentum and the incompressibility constraint leads to the following set of governing equations of finite elastostatics
\begin{equation} \label{Gov-Eq}
\left\{\begin{array}{ll}
{\rm Div}\left[\dfrac{\partial W}{\partial\bfF}(\bfX,\nabla\bfy)-p\nabla\bfy^{-T}\right]=\textbf{0}, & \bfX\in\Omega_0\vspace{0.2cm}\\
\det\nabla\bfy=1, & \bfX\in\Omega_0\vspace{0.2cm}\\
\bfy(\bfX)=\overline{\bfy}(\bfX), & \bfX\in\partial\Omega_0\vspace{0.2cm}\\
\end{array}\right.
\end{equation}
for the deformation field $\bfy(\bfX)$ and the pressure field $p(\bfX)$; note that balance of angular momentum is automatically satisfied thanks to the objectivity of the store-energy functions (\ref{I1-based}).

\subsection{The homogenization limit}

In the limit of separation of length scales between the size of the microstructure and the macroscopic size of the domain $\Omega_0$ occupied by the blend, the solution of (\ref{Gov-Eq}) for the deformation field $\bfy(\bfX)$ and the pressure field $p(\bfX)$ is expected to converge to the solution of the finite elastostatics problem
\begin{equation} \label{Gov-Eq-Hom}
\left\{\begin{array}{ll}
{\rm Div}\left[\dfrac{\partial \overline{W}}{\partial\bfF}(\nabla\bfy)-p\nabla\bfy^{-T}\right]=\textbf{0}, & \bfX\in\Omega_0\vspace{0.2cm}\\
\det\nabla\bfy=1, & \bfX\in\Omega_0\vspace{0.2cm}\\
\bfy(\bfX)=\overline{\bfy}(\bfX), & \bfX\in\partial\Omega_0\vspace{0.2cm}\\
\end{array}\right.
\end{equation}
for a \emph{homogeneous} nonlinear elastic solid with effective stored-energy function $\overline{W}(\bfF)$. What is more, based on the classical homogenization result of Braides \cite{Braides85} and M\"uller \cite{Muller87}, absent geometric instabilities, the expectation is that the effective stored-energy function $\overline{W}(\bfF)$ in (\ref{Gov-Eq-Hom}) is given by the formula
\begin{equation} \label{Weff}
\overline{W}(\obfF)=\displaystyle\int_{\mathcal{Y}_0}W(\bfX,\obfF+\nabla\bfu)\,{\rm d}\bfX,
\end{equation}
where $\obfF$ is any second-order tensor of choice subject to the macroscopic incompressibility constraint $\det\obfF=1$, while the field $\bfu(\bfX)$ is the $\mathcal{Y}_0$-periodic function that, together with the $\mathcal{Y}_0$-periodic pressure field $q(\bfX)$, is defined implicitly as the solution of the \emph{unit-cell problem}
\begin{equation} \label{Gov-Eq-Unit-cell}
\left\{\begin{array}{ll}
{\rm Div}\left[\dfrac{\partial W}{\partial\bfF}(\bfX,\obfF+\nabla\bfu)-q(\obfF+\nabla\bfu)^{-T}\right]=\textbf{0}, & \bfX\in\mathcal{Y}_0\vspace{0.2cm}\\
\det(\obfF+\nabla\bfu)=1, & \bfX\in\mathcal{Y}_0
\end{array}\right. .
\end{equation}

In general, the unit-cell problem (\ref{Gov-Eq-Unit-cell}) does not admit analytical solutions. By now, nevertheless, it is straightforward to solve it numerically by means of a hybrid FE method; see, e.g., \cite{lopez2013nonlinear,LLP17,LWLPN2020}. In this work, we make use of the commercial code ABAQUS to generate FE solutions for (\ref{Gov-Eq-Unit-cell}). In particular, we make use of the hybrid quadratic tetrahedral elements C3D10H built in that code. The conversion of the linear simplicial meshes generated for the unit cells in the previous section to the quadratic simplicial meshes used to solve the equations (\ref{Gov-Eq-Unit-cell}) is carried out in Gmsh \cite{Gmsh09}.

\subsubsection{The limit of small deformations}

In the limit of small deformations as $\obfF\rightarrow\bfI$, the effective stored-energy function (\ref{Weff}) can be shown to reduce to the quadratic form
\begin{equation*} %\label{Weff-Lin}
\overline{W}(\obfF)=\dfrac{1}{2}(\obfF-\bfI)\cdot\overline{\bfL}(\obfF-\bfI)
\end{equation*}
to leading order, where $\obfF$ is now any second-order tensor of choice subject to the linearized macroscopic incompressibility constraint ${\rm tr}\,(\obfF-\bfI)=0$. In this expression, the fourth-order tensor $\overline{\bfL}$, which is nothing more than the effective initial modulus of elasticity of the blend, is given by the formula
\begin{align}
\overline{L}_{ijkl}=\displaystyle\int_{\mathcal{Y}_0} & \left\{2\mu(\bfX)\mathcal{K}_{ijmn}\left(\delta_{mk}\delta_{nl}+\right.\right.\nonumber\\
&\left.\left.\dfrac{\partial \omega_{mkl}}{\partial X_n}(\bfX)\right)+\delta_{ij}\Sigma_{kl}(\bfX)\right\}{\rm d}\bfX, \label{Leff-0}
\end{align}
where we have made use of the notation
\begin{align*}
\mu(\bfX)=\theta_0^{(1)}(\bfX)\mu^{(1)}+\theta_0^{(2)}(\bfX)\mu^{(2)}
\end{align*}
to denote the pointwise initial shear modulus of the blend, $\mathcal{K}_{ijkl}=1/2(\delta_{ik}\delta_{jl}+\delta_{il}\delta_{jk})-1/3\delta_{ij}\delta_{kl}$ is the classical deviatoric projection tensor, and $\omega_{mkl}(\bfX)$ and $\Sigma_{kl}(\bfX)$ are the $\mathcal{Y}_0$-periodic functions defined as the solution of the unit-cell problem
\begin{align}
\left\{\begin{array}{l}
\dfrac{\partial}{\partial X_j}\left[2\mu(\bfX)\mathcal{K}_{ijkl}\dfrac{\partial \omega_{kmn}}{\partial X_l}(\bfX)+\delta_{ij}\Sigma_{mn}(\bfX)\right]=\\
\hspace{1cm}-\dfrac{\partial}{\partial X_j}\left[2\mu(\bfX)\right]\mathcal{K}_{ijmn},\hspace{1.8cm} \qquad \bfX\in \mathcal{Y}_0\vspace{0.4cm}\\
\dfrac{\partial\omega_{imn}}{\partial X_i}(\bfX)=0,\hspace{4.5cm}\bfX\in \mathcal{Y}_0
\end{array}\right. .\label{Eq-omega}
\end{align}

The computation of the effective initial modulus of elasticity (\ref{Leff-0}) for a given blend amounts thus to solving the unit-cell problem (\ref{Eq-omega}) for the functions $\omega_{mkl}(\bfX)$ and $\Sigma_{kl}(\bfX)$ and then carrying out the integral in (\ref{Leff-0}). In general, the unit-cell problem  (\ref{Eq-omega}) can only be solved numerically. By now, again, it is straightforward to do so by means of a hybrid FE method; see, e.g., \cite{LLP17a}.

In Section \ref{Sec: Results 1} below, we present FE results for the effective initial modulus of elasticity (\ref{Leff-0}) of blends for the three values (\ref{H-values}) of the microstructural parameter $H$ and a range of rubber heterogeneity contrasts $t=\mu^{(2)}/\mu^{(1)}$. Since the blends are only approximately isotropic, we present the results in terms of the effective shear modulus $\overline{\mu}$ defined by the isotropic projection
\begin{align}\label{mueff}
\overline{\mu}:=\dfrac{1}{10}\Ktan\cdot\overline{\bfL}=\dfrac{1}{10}\mathcal{K}_{ijkl}\overline{L}_{ijkl}.
\end{align}
In the same section, we also present results for the constitutive anisotropy measure
\begin{align}\label{Ae}
\mathcal{A}_c=\dfrac{||\Ktan\overline{\bfL}\Ktan-2\overline{\mu}\,\Ktan||_{\infty}}{||\Ktan\overline{\bfL}\Ktan||_{\infty}},
\end{align}
which complements the geometric anisotropy measure (\ref{isotropic-filter}) to probe the overall isotropy of the rubber blends; in this last expression, $||\cdot||_{\infty}$ stands for the $L^{\infty}$ norm.

\subsubsection{The special case of Neo-Hookean rubber phases}\label{Sec: NH}

For a given local stored-energy function $W(\bfX,\bfF)$, the resulting effective stored-energy function $\overline{W}(\obfF)$ is, in general, functionally very different from $W(\bfX,\bfF)$. Based on several analytical and computational results worked out over the past decade \cite{lopez2013nonlineara,lopez2013nonlinear,GSPLP15,LLP17a,SLLP19,SPLP20}, the case of a blend made of Neo-Hookean rubber phases is expected to be a rare exception to this rule. Indeed, the effective stored-energy function of any isotropic incompressible composite material made of Neo-Hookean phases is expected to be approximately\footnote{In two space dimensions, it is expected to be \emph{exactly} Neo-Hookean \cite{lefevre2022curious}.} Neo-Hookean.

Precisely, for the problem at hand here, when the nonlinear elastic behaviors of the two rubber phases in the blend are characterized by the Neo-Hookean stored-energy functions (\ref{W-Neo}), the expectation is that the effective stored-energy function (\ref{Weff}) is approximately given by the Neo-Hookean formula
\begin{align}\label{Weff-NH}
\overline{W}(\obfF)=\dfrac{\overline{\mu}}{2}\left[\,\overline{I}_1-3\right],
\end{align}
where we have made use of the notation $\overline{I}_1={\rm tr}\,\overline{\bfC}=\obfF\cdot\obfF$, with $\overline{\bfC}=\obfF^T\obfF$, and where $\overline{\mu}$ stands for the effective initial shear modulus (\ref{mueff}) of the blend. In other words,  the effective stored-energy function (\ref{Weff}) is expected\footnote{Note that for the class of isotropic incompressible blends of interest here, the effective stored-energy function $\overline{W}(\obfF)$ is at most a nonlinear function of the first  and second  macroscopic principal invariants $\overline{I}_1$ and $\overline{I}_2$ of $\overline{\bfC}$.} to depend roughly linearly on $\overline{I}_1$ and to be roughly independent of $\overline{I}_2=1/2[\oI_1^2-{\rm tr}\,\overline{\bfC}^2]$. The results presented in Section \ref{Sec: Results 2} below show that this is indeed the case.

\section{Results in the limit of small deformations}\label{Sec: Results 1}

\begin{figure}[b!]
\centering
\includegraphics[width=.37\textwidth]{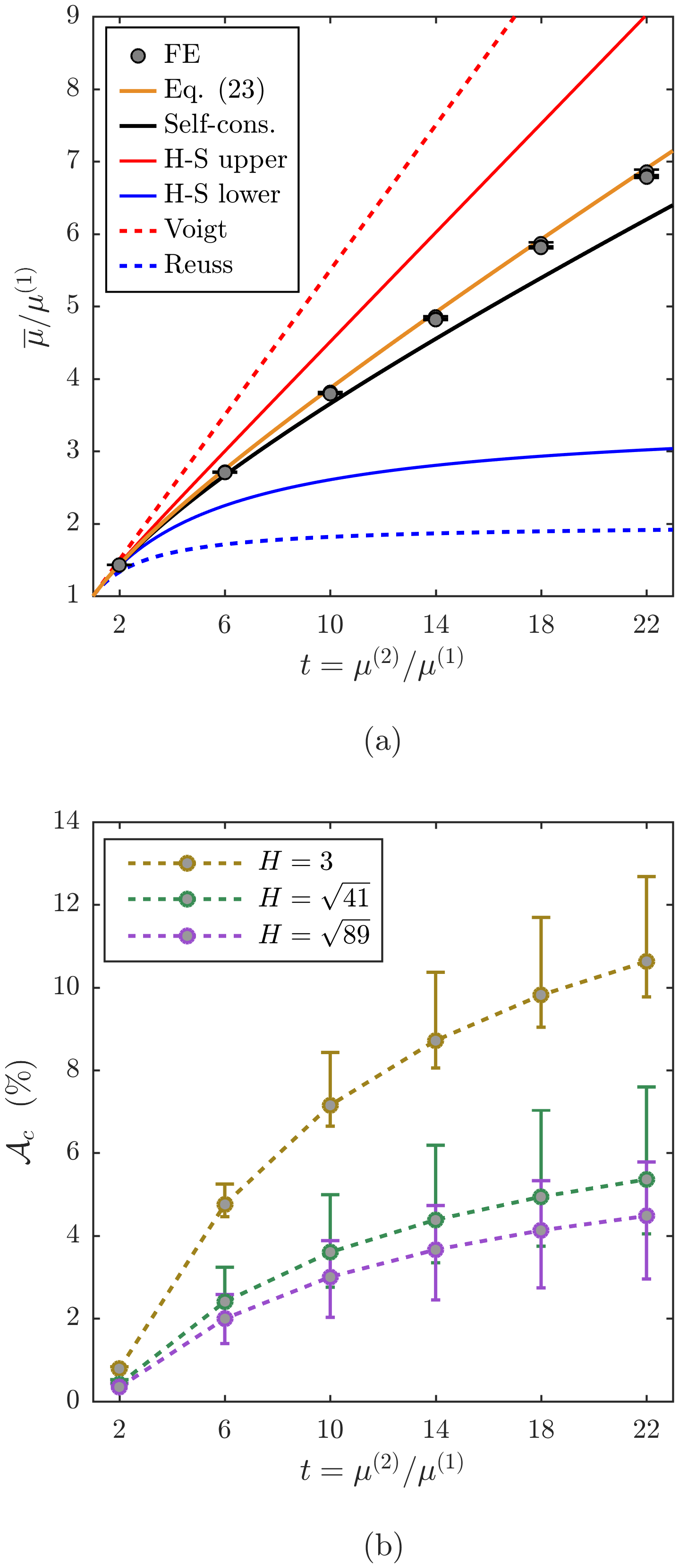}
\caption{(a) FE results for the effective initial shear modulus (\ref{mueff}) of rubber blends, normalized by the initial shear modulus $\mu^{(1)}$ of the rubber phase $r=1$, as a function of the heterogeneity contrast $t=\mu^{(2)}/\mu^{(1)}$ between the two rubber phases. (b) The constitutive anisotropy measure (\ref{Ae}) of the same blends, plotted also as a function of $t$. The results correspond to 30 realizations of unit cells, 10 for each of the values (\ref{H-values}) for the microstructural parameter $H$. For direct comparison, the formula (\ref{mueff-formula}), the self-consistent estimate (\ref{mueffSC}), the Hashin-Shtrikman bounds (\ref{mueffHS}), and the Voigt and Reuss bounds (\ref{mueffVR}) are also included in part (a).}
\label{Fig7}
\end{figure}

Figure \ref{Fig7}(a) presents FE results for the effective initial shear modulus $\overline{\mu}/\mu^{(1)}$, normalized by the initial shear modulus of the rubber phase $r=1$, as a function of the heterogeneity contrast $t=\mu^{(2)}/\mu^{(1)}$ between the two rubber phases. The results correspond to contrasts $t=2,6,10,14,18,22$ for a total of 30 realizations, 10 unit cells for each of the three values $H=3, \sqrt{41}$, and $\sqrt{89}$. To aid the analysis of the results, the figure also includes plots of the fitting formula\footnote{The formula (\ref{mueff-formula}) is nothing more than a linear interpolation between the self-consistent estimate (\ref{mueffSC}) and the HS upper bound (\ref{mueffHS})$_1$.}
\begin{equation}\label{mueff-formula}
\overline{\mu}=\dfrac{1}{16}\left(\dfrac{1+19t+10t^2}{1+4t}+\sqrt{1+98t+t^2}\right)\mu^{(1)}
\end{equation}
of the FE results, the self-consistent estimate \cite{Hill65,Budiansky65,Willis77}
\begin{equation}\label{mueffSC}
\overline{\mu}^{\,{\rm sc}}=\dfrac{1}{12}\left(1+t+\sqrt{1+98t+t^2}\right)\mu^{(1)},
\end{equation}
the Hashin-Shtrikman (HS) upper and lower bounds \cite{HS63}
\begin{equation}\label{mueffHS}
\left\{\begin{array}{l}
\overline{\mu}^{\,{\rm HS},u}=\left(1-\dfrac{(1-t)(2+3t)}{2(1+4t)}\right)\mu^{(1)}\vspace{0.2cm}\\
\overline{\mu}^{\,{\rm HS},l}=\left(1+\dfrac{5(t-1)}{2(4+t)}\right)\mu^{(1)}\end{array}\right. ,
\end{equation}
and the Voigt \cite{Voigt89} and Reuss \cite{Reuss29} bounds
\begin{equation}\label{mueffVR}
\left\{\begin{array}{l}
\overline{\mu}^{\,{\rm V}}=\dfrac{1}{2}(1+t)\mu^{(1)}\vspace{0.2cm}\\
\overline{\mu}^{\,{\rm R}}=\dfrac{2t}{1+t}\mu^{(1)}\end{array}\right. .
\end{equation}
\begin{remark}
Bicontinuous rubber blends used in applications rarely exceed a heterogeneity contrast of $t=10$, which is well within the range of values $t\in[1,22]$ considered here.
\end{remark}

Remarkably, the results in Fig. \ref{Fig7}(a) show that the initial elastic response of the blends is largely insensitive to the details of their microstructure, at least for the range of heterogeneity contrasts $t\in[1,22]$ considered here. Indeed, the computed values for effective initial shear modulus $\overline{\mu}$ show little difference for different realizations and for the different values of $H$, which, again, is a parameter that is directly related to the characteristic length scale of the microstructure within the unit cell. The results also show that the effective initial shear modulus $\overline{\mu}$ is well within the HS bounds and that it is reasonably well approximated by the self-consistent estimate (\ref{mueffSC}). Nevertheless, one can do better, as the formula (\ref{mueff-formula}) makes it plain.

The results in Fig. \ref{Fig7}(b) provide a quantification of the extent to which rubber blends with microstructures based on different values for the microstructural parameter $H$, and different realizations for the same value of $H$, deviate from exact isotropy. As expected from their construction process, rubber blends with larger $H$ are seen to be more isotropic and different realizations exhibit a range of deviations from isotropy. As also expected from basic physical intuition, the deviation from exact isotropy is seen to increase with increasing heterogeneity contrast between the two rubber phases.

\section{Results for blends of Neo-Hookean rubbers}\label{Sec: Results 2}

Next, we turn to the FE results for the special case when the nonlinear elastic behaviors of both rubber phases in the blend are characterized by the Neo-Hookean stored-energy functions (\ref{W-Neo}). For clarity of presentation, throughout this section, we restrict attention to a rubber blend with microstructure based on the microstructural parameter $H=3$ and the heterogeneity contrast $t=\mu^{(2)}/\mu^{(1)}=10$ between the two Neo-Hookean rubber phases. The conclusions remain fundamentally the same for the other two values $H=\sqrt{41}$ and  $\sqrt{89}$ of the parameter $H$ and all the other 5 heterogeneity contrasts in the range $t\in[1,22]$ that we have examined.

\begin{remark}\label{R: contrast}
Note that in the context of finite elasticity, the actual heterogeneity contrast is \emph{not} constant since the two rubber phases undergo very different deformations during loading. Nevertheless, we find the labeling of $t=\mu^{(2)}/\mu^{(1)}$ as the ``heterogeneity contrast'' convenient for the special case of Neo-Hookean rubber blends.
\end{remark}

\begin{figure}[t!]
\centering
\includegraphics[width=.37\textwidth]{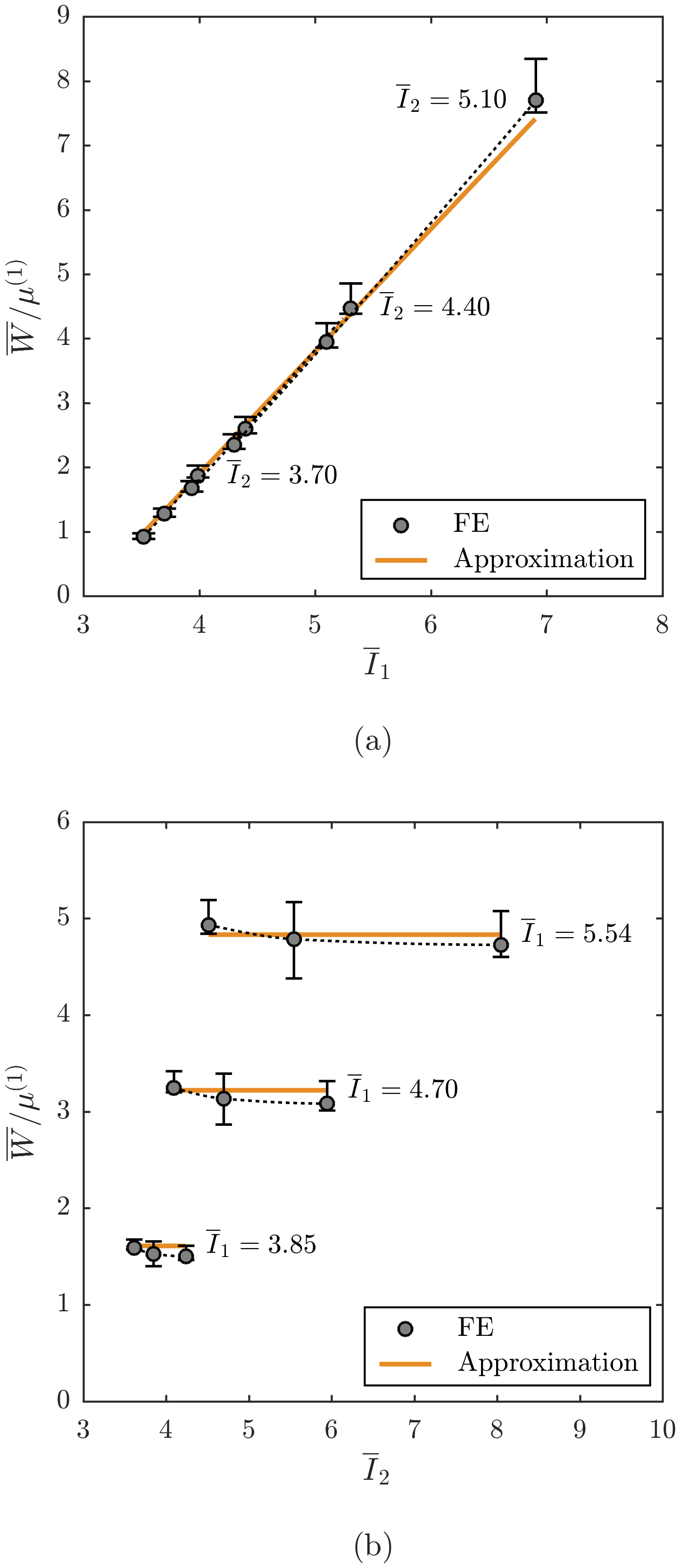}
\caption{FE results for the effective stored-energy function (\ref{Weff}) for a blend of Neo-Hookean rubber phases plotted as a function of each of the two macroscopic invariants $\oI_1$ and $\oI_2$ for three fixed values of the remaining invariant. The results correspond to 10 realizations of unit cells, all of them generated with the microstructural parameter $H=3$, the  heterogeneity contrast $t=\mu^{(2)}/\mu^{(1)}=10$, and are shown normalized by the initial shear modulus $\mu^{(1)}$ of the rubber phase $r=1$. The solid lines correspond to the approximation (\ref{Weff-NH}) with (\ref{mueff-formula}).}
\label{Fig8}
\end{figure}

In light of the expectation discussed in Subsection \ref{Sec: NH} above that the macroscopic elastic response of Neo-Hookean rubber blends is approximately Neo-Hookean, we begin by presenting results for the effective stored-energy function $\overline{W}$ as a function of each of the two macroscopic invariants $\oI_1$ and $\oI_2$, while keeping the remaining invariant fixed. Figure \ref{Fig8} presents such results for three representative cases. Part (a) of the figure shows $\overline{W}/\mu^{(1)}$ as a function of
$\overline{I}_1$ for the fixed values $\oI_2=3.70, 4.40,$ and $5.10$ of the second invariant. Similarly, part (b) shows $\overline{W}/\mu^{(1)}$ as a function of
$\overline{I}_2$ for the fixed values $\oI_1=3.85, 4.70,$ and $5.54$ of the first invariant. Here, it is important to remark that fixing the value of one of the invariants $\oI_1$ or $\oI_2$ restricts the range of physically allowable values of the remaining invariant. The results displayed in Fig. \ref{Fig8}, which correspond to 10 different realizations of unit cells, pertain to the entire range of allowable values for each of the cases that is presented.

\begin{figure*}[t!]
\centering
\includegraphics[width=\textwidth]{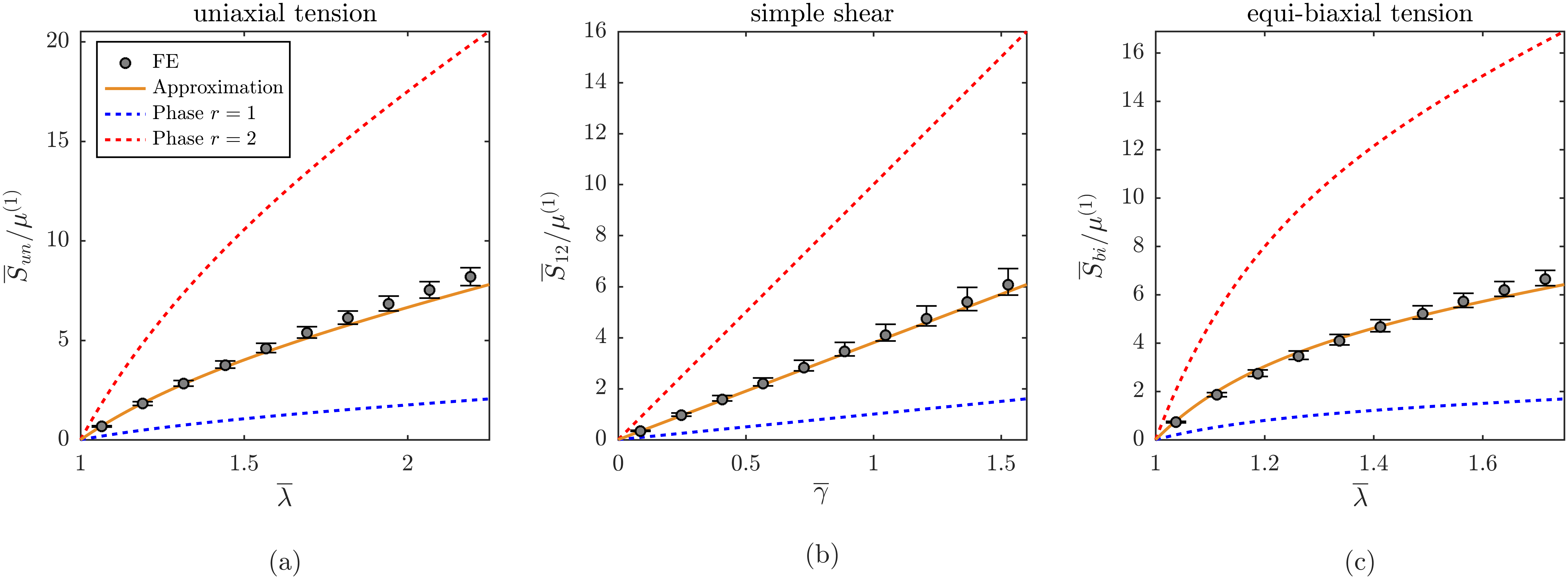}
\caption{FE results for the macroscopic stress-deformation response (\ref{S-F-macro}) of a blend of Neo-Hookean rubber phases under: (a) uniaxial tension, (b) simple shear, and (c) equi-biaxial tension. The results correspond to 10 realizations of unit cells, all of them generated with the microstructural parameter $H=3$, and the heterogeneity contrast $t=\mu^{(2)}/\mu^{(1)}=10$. For direct comparison, the corresponding responses (\ref{S-F-Approx}) predicted by the approximation (\ref{Weff-NH}) with (\ref{mueff-formula}), as well as those of the underlying rubber phases $r=1$ and 2, are also plotted.}
\label{Fig9}
\end{figure*}
\begin{figure*}[t!]
\centering
\includegraphics[width=\textwidth]{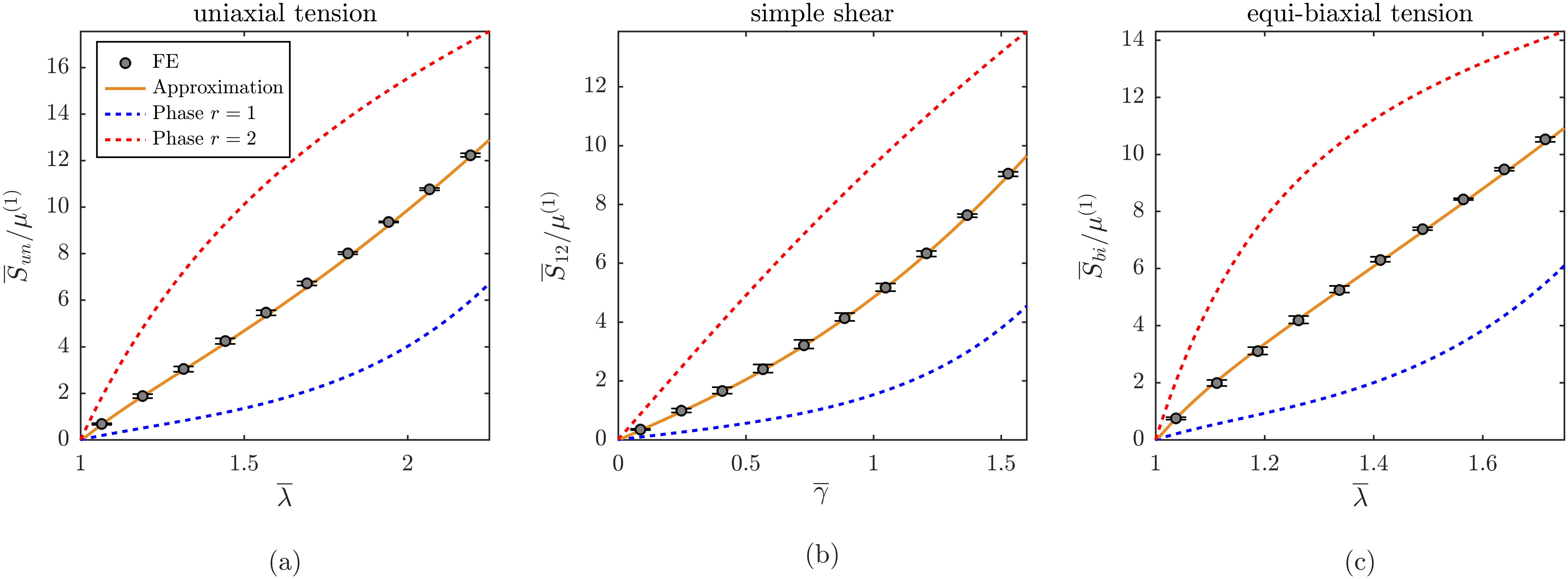}
\\
\vspace{.1in}
\includegraphics[width=\textwidth]{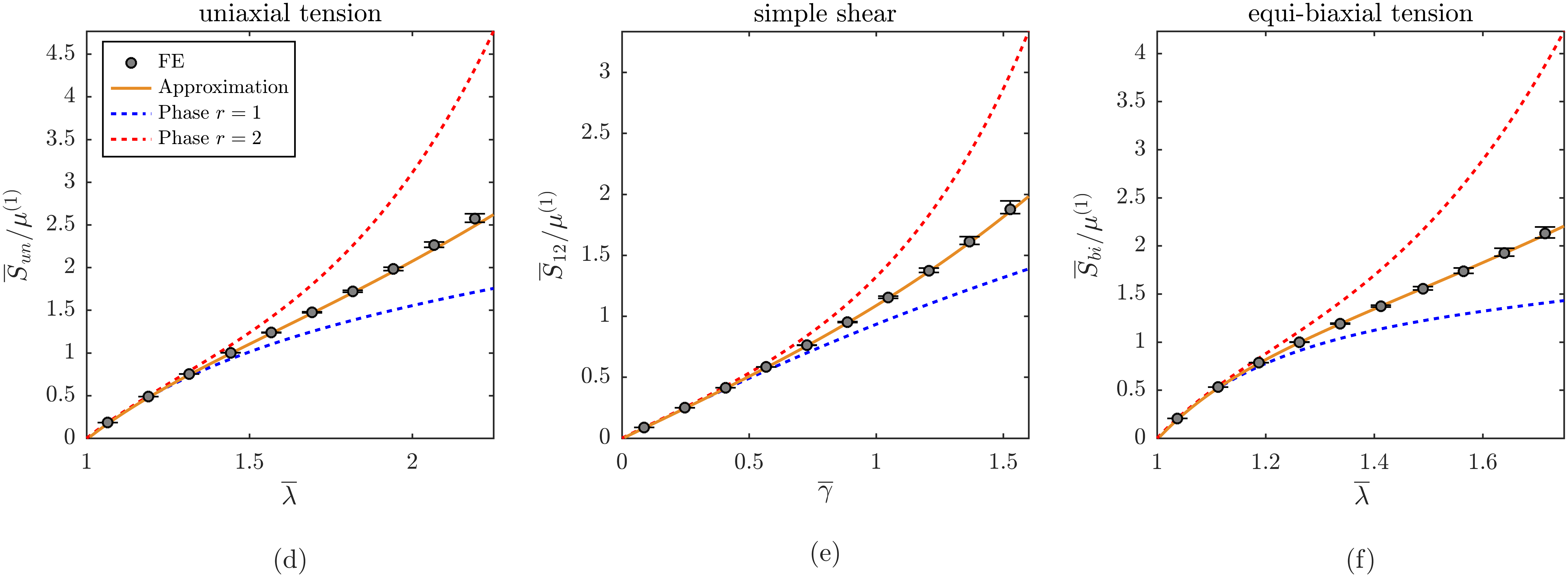}
\vspace{-.25in}
\caption{FE results for the macroscopic stress-deformation response (\ref{S-F-macro}) of blends of non-Gaussian rubbers with stored-energy functions (\ref{two-term}) under: (a,d) uniaxial tension, (b,e) simple shear, and (c,f) equi-biaxial tension. The results correspond to Blends I (a,b,c) and II (d,e,f), with the material constants listed in Table \ref{Table1}, and 10 realizations of unit cells, all of them generated with the microstructural parameter $H=3$. For direct comparison, the corresponding responses predicted by the approximation (\ref{Weff-Approx-CM})-(\ref{Equations-mu0}), as well as those of the underlying rubber phases $r=1$ and 2, are also plotted.}
\label{Fig10}
\end{figure*}

Three observations are immediate from Fig. \ref{Fig8}. The effective stored-energy function $\overline{W}$ for the Neo-Hookean rubber blend depends roughly linearly on $\oI_1$, is essentially independent of $\oI_2$, and is well approximated by the fully explicit formula (\ref{Weff-NH}) with (\ref{mueff-formula}). These observations fall squarely within the expectations discussed in Subsection \ref{Sec: NH}.

To gain further insight into the response of the Neo-Hookean rubber blend, Fig. \ref{Fig9} presents results for its stress-deformation response
\begin{equation}\label{S-F-macro}
\obfS=\dfrac{\partial \overline{W}}{\partial\obfF}(\obfF)-p \obfF^{-T}
\end{equation}
under:
\begin{itemize}[itemsep=1pt]

\item{Uniaxial tension when $\obfF=\ol\bfe_1\otimes\bfe_1+\ol^{-1/2}(\bfe_2\otimes\bfe_2+\bfe_3\otimes\bfe_3)$ with $\obfS=\overline{S}_{un}\bfe_1\otimes\bfe_1$ and prescribed $\ol\geq 1$;}

\item{Simple shear when $\obfF=\bfI+\overline{\gamma}\,\bfe_1\otimes\bfe_2$ with $\obfS=\overline{S}_{11}\bfe_1\otimes\bfe_1+\overline{S}_{12}\bfe_1\otimes\bfe_2+\overline{S}_{21}\bfe_2\otimes\bfe_1+\overline{S}_{22}\bfe_2\otimes\bfe_2$ and prescribed $\overline{\gamma}\geq 0$; and}

\item{Equi-biaxial tension when $\obfF=\ol^{\,-2}\bfe_1\otimes\bfe_1+\ol(\bfe_2\otimes\bfe_2+\bfe_3\otimes\bfe_3)$ with $\obfS=\overline{S}_{bi}(\bfe_2\otimes\bfe_2+\bfe_3\otimes\bfe_3)$ and prescribed $\ol\geq 1$.}

\end{itemize}
The results pertain to the same 10 realizations of unit cells presented in Fig. \ref{Fig8}. For direct comparison, the figure includes the corresponding responses of the two rubber phases $r=1$ and 2, as well as those predicted by the approximation (\ref{Weff-NH}) with (\ref{mueff-formula}), which read
\begin{equation}\label{S-F-Approx}
\overline{S}_{un}=\overline{\mu}\left(\ol-\ol^{\,-2}\right),\quad \overline{S}_{12}=\overline{\mu}\,\overline{\gamma},\quad
\overline{S}_{bi}=\overline{\mu}\left(\ol-\ol^{\,-5}\right).
\end{equation}

A quick glance at Fig. \ref{Fig9} suffices to recognize that the response of the Neo-Hookean blend is indeed approximately described by the Neo-Hookean formula (\ref{Weff-NH}) with (\ref{mueff-formula}). This observation, when combined with the conclusions established in Section \ref{Sec: Results 1} above for the effective initial shear modulus $\overline{\mu}$, implies that the nonlinear elastic response of Neo-Hookean blends is essentially independent of the details of their microstructure, irrespective of the applied deformation. In other words --- rather strikingly --- all 50/50 bicontinuous blends of Neo-Hookean rubbers behave substantially in the same manner, irrespective of the morphologies of their phases.

\section{Results for blends of non-Gaussian rubbers}\label{Sec: Results 3}

Finally, we present FE results for the case that is most often encountered in practice, that of blends made of two different non-Gaussian rubbers. For definiteness, we consider that the elastic behaviors of the two rubber phases are characterized by the stored-energy functions \cite{lopez2010new}
\begin{equation} \label{two-term}
\psi^{(r)}(I_1) = \sum_{s=1}^2 \dfrac{3^{1-\alpha^{(r)}_s}}{2 \alpha_s }\mu^{(r)}_s \left[I_1^{\alpha^{(r)}_s} - 3^{\alpha^{(r)}_s} \right]
\end{equation}
$(r=1,2)$, where $\mu^{(r)}_s$ and $\alpha^{(r)}_s$ are real-valued material constants. In view of the linearization and ellipticity conditions (\ref{linarization-cond})-(\ref{strong-ell-cond}), note that $\mu^{(r)}_1, \mu^{(r)}_2>0$, $\mu^{(r)}_1+\mu^{(r)}_2=\mu^{(r)}$, and that $\max_{s}\{\alpha^{(r)}_s\}$ must be positive and sufficiently large.

\begin{table*}[t!]\centering
\caption{Material constants in the stored-energy functions (\ref{two-term}) for the two types of blends of non-Gaussian rubbers considered in the simulations.}
\begin{tabular}{l|cccc|cccc|c}
\toprule
Blend I & $\mu^{(1)}_1$ & $\mu^{(1)}_2$ & $\alpha^{(1)}_1$  & $\alpha^{(1)}_2$ & $\mu^{(2)}_1$ & $\mu^{(2)}_2$ & $\alpha^{(2)}_1$  & $\alpha^{(2)}_2$ & Initial contrast $t=\mu^{(2)}/\mu^{(1)}$\\
\midrule
       & $0.5$ & $0.5$ & $3.5$ & $1$ & $5$ & $5$ & $1$ & $0.5$& $10$\\
\midrule
\midrule
Blend II & $\mu^{(1)}_1$ & $\mu^{(1)}_2$ & $\alpha^{(1)}_1$  & $\alpha^{(1)}_2$ & $\mu^{(2)}_1$ & $\mu^{(2)}_2$ & $\alpha^{(2)}_1$  & $\alpha^{(2)}_2$& Initial contrast $t=\mu^{(2)}/\mu^{(1)}$\\
\midrule
       & $0.5$ & $0.5$ & $1$ & $0.5$ & $0.5$ & $0.5$ & $3$ & $0.5$ & 1\\
\bottomrule
\end{tabular} \label{Table1}
\end{table*}

Now, in the context of finite deformations,  as already noted in Remark \ref{R: contrast} above, the heterogeneity contrast between the two rubber phases in a blend is \emph{not} fixed but evolves with deformation. For this reason, we consider two types of blends, one --- labeled Blend I --- wherein the initial contrast is large but the initially softer rubber phase $(r=1)$ stiffens faster than the initially stiffer rubber phase $(r=2)$ and, viceversa, one --- labeled Blend II ---  wherein the initial contrast is small but one of the rubber phases $(r=2)$ stiffens faster than the other rubber phase $(r=1)$. In particular, we consider blends of the non-Gaussian rubbers with the stored-energy functions (\ref{two-term}) and the two sets of material constants listed in Table \ref{Table1}.

Figure \ref{Fig10} presents results for the stress-deformation responses (\ref{S-F-macro}) of the two types of blends, Blends I and II. The results correspond to the same 10 different realizations of unit cells based on the value $H=3$ for the microstructural parameter $H$. The results for the other two values $H=\sqrt{41}$ and $\sqrt{89}$
are essentially identical. Moreover, the results correspond to the same three loadings considered above for the Neo-Hookean blends, to wit, uniaxial tension, simple shear, and equi-biaxial tension. For direct comparison, the corresponding responses of the two rubber phases $r = 1$ and $2$, as well as the predictions generated by the analytical approximation (\ref{Weff-Approx-CM})-(\ref{Equations-mu0}) introduced in Section \ref{Sec: Approximation} below, are also plotted in the figures.

Four observations are immediate from Fig. \ref{Fig10}. First, as expected, the macroscopic response of the blend is some sort of nonlinear combination of the responses of its rubber phases. Second, the difference in the stress-deformation response among different realizations of blends of non-Gaussian rubbers is noticeably smaller than that of blends of Neo-Hookean rubbers. This is because the evolution of the heterogeneity contrast in the non-Gaussian blends leads to effectively smaller contrasts than in the Neo-Hookean blends. Third, the analytical approximation (\ref{Weff-Approx-CM})-(\ref{Equations-mu0}) is in good agreement with the FE results. Finally, consistent with the previous observations in the limit of small observations and for Neo-Hookean blends, the nonlinear elastic response of blends of non-Gaussian rubbers is also largely insensitive to the morphologies of the underlying rubber phases.

\section{An analytical approximation}\label{Sec: Approximation}

In the sequel, complementary to the FE results presented in the three preceding sections, we work out an analytical approximation for the effective stored-energy function (\ref{Weff}). We do so by making use of the nonlinear comparison medium method\footnote{The interested reader is referred to Sections 1 and 4 in \cite{lopez2013nonlinear} for an account of the historical development of comparison medium methods, whose origins date back to the celebrated work of Talbot and Willis \cite{talbot1985variational}.} introduced in \cite{lopez2013nonlinear}, together with the approximate solution (\ref{Weff-NH}) for the effective stored-energy function of Neo-Hookean rubber blends.

We begin by defining the local stored-energy function of a ``comparison rubber blend'' with the same bicontinuous microstructure as the rubber blend of interest, that is, with the same characteristic functions $\theta_0^{(1)}(\bfX)$ and $\theta_0^{(2)}(\bfX)$ as in (\ref{indicator-functions}), but possibly different isotropic incompressible rubber phases. Analogous to (\ref{W-local}), we write
\begin{equation*} %\label{I1-based-blend}
W_0(\bfX,\bfF)=\theta_0^{(1)}(\bfX)\psi_0^{(1)}(I_1)+\theta_0^{(2)}(\bfX)\psi_0^{(2)}(I_1).
\end{equation*}
As elaborated in Section 4 in \cite{lopez2013nonlinear}, it follows that the effective stored-energy function (\ref{Weff}) of any rubber blend of interest can be approximated variationally directly in terms of the effective stored-energy function
\begin{equation} \label{W0eff}
\overline{W}_0(\obfF)=\displaystyle\min_{\bfy\in \mathcal{K}_a}\displaystyle\int_{\mathcal{Y}_0}W_0(\bfX,\obfF+\nabla\bfu)\,{\rm d}\bfX
\end{equation}
of the ``comparison rubber blend''; in this last expression,  $\mathcal{K}_a$ stands for the set of kinematically admissible deformation fields of the form $\bfy(\bfX)=\obfF\bfX+\bfu(\bfX)$, where $\bfu(\bfX)$ is $\mathcal{Y}_0$-periodic, that satisfy the incompressibility constraint $\det\nabla\bfy=1$. The result reads
\begin{align*}
\overline{W}(\obfF)\geq & \overline{W}_0(\obfF)+c_0^{(1)}\min_{\mathcal{I}_1^{(1)}}\left\{\psi^{(1)}(\mathcal{I}_1^{(1)})-\psi_0^{(1)}(\mathcal{I}_1^{(1)})\right\}+
\vspace{0.1cm}\\
& c_0^{(2)}\min_{\mathcal{I}_1^{(2)}}\left\{\psi^{(2)}(\mathcal{I}_1^{(2)})-\psi_0^{(2)}(\mathcal{I}_1^{(2)})\right\}
\end{align*}
and
\begin{align*}
\overline{W}(\obfF)\leq & \overline{W}_0(\obfF)+c_0^{(1)}\max_{\mathcal{I}_1^{(1)}}\left\{\psi^{(1)}(\mathcal{I}_1^{(1)})-\psi_0^{(1)}(\mathcal{I}_1^{(1)})\right\}+
\vspace{0.1cm}\\
& c_0^{(2)}\max_{\mathcal{I}_1^{(2)}}\left\{\psi^{(2)}(\mathcal{I}_1^{(2)})-\psi_0^{(2)}(\mathcal{I}_1^{(2)})\right\},
\end{align*}
where we emphasize that the stored-energy functions $\psi_0^{(1)}(I_1)$ and $\psi_0^{(2)}(I_1)$ describing the elasticity of the rubber phases in the ``comparison rubber blend'' are, at this stage, arbitrary.

Now, by choosing the rubber phases in the ``comparison rubber blend'' to be Neo-Hookean with stored-energy functions
\begin{align}\label{W0-Neo}
\psi_0^{(1)}(I_1)=\dfrac{\mu_0^{(r)}}{2}[I_1-3]
\end{align}
$(r=1,2)$ and, in turn, by approximating the resulting effective stored-energy function (\ref{W0eff}) with the result (\ref{Weff-NH}), the above inequalities specialize to
\begin{align}
\left\{\begin{array}{l}\label{Weff-CM-0}
\overline{W}(\obfF)\geq \displaystyle\min_{\mathcal{I}_1^{(1)},\,\mathcal{I}_1^{(2)}}\mathcal{W}\left(\oI_1;\mu_0^{(1)},\mu_0^{(2)},\mathcal{I}_1^{(1)},\mathcal{I}_1^{(2)}\right)\\
\rm{and} \vspace{0.2cm}\\
\overline{W}(\obfF)\leq \displaystyle\max_{\mathcal{I}_1^{(1)},\,\mathcal{I}_1^{(2)}}\mathcal{W}\left(\oI_1;\mu_0^{(1)},\mu_0^{(2)},\mathcal{I}_1^{(1)},\mathcal{I}_1^{(2)}\right)\end{array}\right.
\end{align}
with
\begin{align*}
\mathcal{W}:=&\dfrac{\overline{\mu}_0}{2}\left[\,\oI_1-3\right]+c_0^{(1)}\left\{\psi^{(1)}(\mathcal{I}_1^{(1)})-\dfrac{\mu_0^{(1)}}{2}\left[\mathcal{I}_1^{(1)}-3\right]\right\}
+\\
&c_0^{(2)}\left\{\psi^{(2)}(\mathcal{I}_1^{(2)})-\dfrac{\mu_0^{(2)}}{2}\left[\mathcal{I}_1^{(2)}-3\right]\right\},
\end{align*}
where $\overline{\mu}_0=\overline{\mu}_0(\mu_0^{(1)},\mu_0^{(2)})$ is the effective initial shear modulus of the ``comparison rubber blend'' and where we emphasize that the inequalities (\ref{Weff-CM-0}) are valid for any choice of the initial shear moduli $\mu_0^{(1)}$ and $\mu_0^{(2)}$ of the rubber phases in the ``comparison rubber blend''. Optimizing with respect to $\mu_0^{(1)}$ and $\mu_0^{(2)}$ then yields
\begin{align}\label{Weff-CM-1}
\left\{\begin{array}{l}\overline{W}(\obfF)\geq \displaystyle\max_{\mu_0^{(1)},\, \mu_0^{(2)}}\displaystyle\min_{\mathcal{I}_1^{(1)},\,\mathcal{I}_1^{(2)}}\mathcal{W}\left(\oI_1;\mu_0^{(1)},\mu_0^{(2)},\mathcal{I}_1^{(1)},\mathcal{I}_1^{(2)}\right)\\
\rm{and} \vspace{0.2cm}\\
\overline{W}(\obfF)\leq \displaystyle\min_{\mu_0^{(1)},\, \mu_0^{(2)}}\displaystyle\max_{\mathcal{I}_1^{(1)},\,\mathcal{I}_1^{(2)}}\mathcal{W}\left(\oI_1;\mu_0^{(1)},\mu_0^{(2)},\mathcal{I}_1^{(1)},\mathcal{I}_1^{(2)}\right)\end{array}\right. .
\end{align}

Note that the variational approximation (\ref{Weff-CM-1})$_1$ is non-trivial only when \emph{both} stored-energy functions $\psi^{(1)}(I_1)$ and $\psi^{(2)}(I_1)$ have stronger growth conditions than (\ref{W0-Neo}). Similarly, the variational approximation (\ref{Weff-CM-1})$_2$ yields a non-trivial result only when \emph{both} stored-energy functions $\psi^{(1)}(I_1)$ and $\psi^{(2)}(I_1)$ have weaker growth conditions than (\ref{W0-Neo}).

\begin{remark}
Because of the finite length of the polymer chains that they are made of, the elasticity of rubbers is necessarily non-Gaussian and hence their stored-energy functions typically exhibit stronger than Neo-Hookean growth. Accordingly, the variational approximation (\ref{Weff-CM-1})$_1$ is the one that applies in practice.
\end{remark}

For the case when the variational approximations (\ref{Weff-CM-1}) are non-trivial, the optimality conditions with respect to $\mathcal{I}_1^{(1)}$ and $\mathcal{I}_1^{(2)}$ are given by
\begin{equation*}%\label{Opt-I1}
\left\{\begin{array}{l}
\dfrac{{\rm d} \psi^{(1)}}{{\rm d} \, \mathcal{I}_1^{(1)}}(\mathcal{I}_1^{(1)})=\dfrac{\mu_0^{(1)}}{2}\vspace{0.2cm}\\
\dfrac{{\rm d}  \psi^{(2)}}{{\rm d} \,\mathcal{I}_1^{(2)}}(\mathcal{I}_1^{(2)})=\dfrac{\mu_0^{(2)}}{2}
\end{array}\right. ,
\end{equation*}
while the optimality conditions with respect to $\mu_0^{(1)}$ and $\mu_0^{(2)}$ read
\begin{equation}\label{Opt-mu0}
\left\{\begin{array}{l}
\dfrac{\partial\,\overline{\mu}_0}{\partial\mu_0^{(1)}}\left[\,\oI_1-3\right]=c_0^{(1)}\left[\,\mathcal{I}_1^{(1)}-3\right]\vspace{0.2cm}\\
\dfrac{\partial\,\overline{\mu}_0}{\partial\mu_0^{(2)}}\left[\,\oI_1-3\right]=c_0^{(2)}\left[\,\mathcal{I}_1^{(2)}-3\right]
\end{array}\right. .
\end{equation}

Upon recognizing that $\overline{\mu}_0=\overline{\mu}_0(\mu_0^{(1)},\mu_0^{(2)})$ is a homogeneous function of degree 1, and hence that
\begin{equation*}
\overline{\mu}_0=\mu_0^{(1)}\dfrac{\partial\, \overline{\mu}_0}{\partial \mu_0^{(1)}}+\mu_0^{(2)}\dfrac{\partial\, \overline{\mu}_0}{\partial \mu_0^{(2)}}
\end{equation*}
from Euler's theorem of homogeneous functions, and making direct use of relations (\ref{Opt-mu0}), it is straightforward to show that the variational approximations (\ref{Weff-CM-1}) reduce to
\begin{equation}\label{Weff-Approx-CM}
\overline{W}(\obfF)=c_0^{(1)}\psi^{(1)}\left(\mathcal{I}_1^{(1)}\right)+c_0^{(2)}\psi^{(2)}\left(\mathcal{I}_1^{(2)}\right)
\end{equation}
with
\begin{equation*}
\left\{\begin{array}{l}
\mathcal{I}_1^{(1)}=\dfrac{1}{c_0^{(1)}}\dfrac{\partial\,\overline{\mu}_0}{\partial\mu_0^{(1)}}\left[\,\oI_1-3\right]+3\vspace{0.2cm}\\
\mathcal{I}_1^{(2)}=\dfrac{1}{c_0^{(2)}}\dfrac{\partial\,\overline{\mu}_0}{\partial\mu_0^{(2)}}\left[\,\oI_1-3\right]+3
\end{array}\right. ,
\end{equation*}
where $\mu_0^{(1)}$ and $\mu_0^{(2)}$ are solutions of the system of nonlinear algebraic equations
\begin{equation}\label{Equations-mu0}
\left\{\begin{array}{l}
\dfrac{{\rm d}  \psi^{(1)}}{{\rm d} \,\mathcal{I}_1^{(1)}}\left(\dfrac{1}{c_0^{(1)}}\dfrac{\partial\,\overline{\mu}_0}{\partial\mu_0^{(1)}}\left[\,\oI_1-3\right]+3\right)=\dfrac{\mu_0^{(1)}}{2}\vspace{0.2cm}\\
\dfrac{{\rm d}  \psi^{(2)}}{{\rm d} \,\mathcal{I}_1^{(2)}}\left(\dfrac{1}{c_0^{(2)}}\dfrac{\partial\,\overline{\mu}_0}{\partial\mu_0^{(2)}}\left[\,\oI_1-3\right]+3\right)=\dfrac{\mu_0^{(2)}}{2}
\end{array}\right.
\end{equation}
and where, for notational simplicity, we have made use of the equals sign in (\ref{Weff-Approx-CM}) to denote a variational approximation.

\begin{remark}
In general, equations (\ref{Equations-mu0}) do not admit analytical solutions and hence must be solved numerically as a function of $\oI_1$. When solving the equations (\ref{Equations-mu0}) numerically, note that $\mu_0^{(1)}=\mu^{(1)}$ and $\mu_0^{(2)}=\mu^{(2)}$ at $\oI_1=3$, by virtue of the linearization conditions (\ref{linarization-cond}). As $\oI_1$ increases from $3$, the values of $\mu_0^{(1)}$ and $\mu_0^{(2)}$ evolve away from the initial shear moduli $\mu^{(1)}$ and $\mu^{(2)}$ of the rubber phases.
\end{remark}

The approximation (\ref{Weff-Approx-CM})-(\ref{Equations-mu0}) is valid for rubber blends with any microstructure. When specialized to the 50/50 bicontinuous rubber blends of interest in this work, $c_0^{(1)}=c_0^{(2)}=1/2$. Moreover, making use of the result (\ref{mueff-formula}),
\begin{align*}
\overline{\mu}_0=&\dfrac{\mu_0^{(1)}}{16}+\dfrac{5 \mu_0^{(1)}\mu_0^{(2)} (3 \mu_0^{(1)}+2 \mu_0^{(2)})}{16\mu_0^{(1)} (\mu_0^{(1)}+4 \mu_0^{(2)})}+\\
&\dfrac{\mu_0^{(1)}}{16}\sqrt{1+98\dfrac{\mu_0^{(2)}}{\mu_0^{(1)}}+\left(\dfrac{\mu_0^{(2)}}{\mu_0^{(1)}}\right)^2},
\end{align*}
and therefore
\begin{equation*}
\left\{\begin{array}{l}
\dfrac{\partial\,\overline{\mu}_0}{\partial\mu_0^{(1)}}=\dfrac{1}{16}+\dfrac{1}{16}\sqrt{1+98\dfrac{\mu_0^{(2)}}{\mu_0^{(1)}}+\left(\dfrac{\mu_0^{(2)}}{\mu_0^{(1)}}\right)^2}-\vspace{0.2cm}\\
\hspace{1.1cm}\dfrac{\left(49+\frac{\mu_0^{(2)}}{\mu_0^{(1)}}\right)\mu_0^{(2)}}{16\mu_0^{(1)}\sqrt{1+98\dfrac{\mu_0^{(2)}}{\mu_0^{(1)}}+\left(\dfrac{\mu_0^{(2)}}{\mu_0^{(1)}}\right)^2}}+\\
\hspace{1.1cm}\dfrac{50\left(\mu_0^{(2)}\right)^2}{16(\mu_0^{(1)}+4\mu_0^{(2)})^2}\vspace{0.2cm}\\
\dfrac{\partial\,\overline{\mu}_0}{\partial\mu_0^{(2)}}=\dfrac{5}{32}+\dfrac{49 \mu_0^{(1)}+\mu_0^{(2)}}{16\mu_0^{(1)} \sqrt{1+98\dfrac{\mu_0^{(2)}}{\mu_0^{(1)}}+\left(\dfrac{\mu_0^{(2)}}{\mu_0^{(1)}}\right)^2}}-\\
\hspace{1.1cm}\dfrac{5 \mu_0^{(2)} (3 \mu_0^{(1)}+2 \mu_0^{(2)})}{4(\mu_0^{(1)}+4 \mu_0^{(2)})^2}+\dfrac{25 \mu_0^{(1)}+20 \mu_0^{(2)}}{32(\mu_0^{(1)}+4 \mu_0^{(2)})}\\
\end{array}\right. .
\end{equation*}

\section{Final comments}\label{Sec:FinalComments}

The results presented in this work have provided a first step in directly linking the macroscopic properties of bicontinuous rubber blends to their microstructure. Rather unexpectedly, the results have revealed that the nonlinear elastic response of isotropic 50/50 bicontinuous rubber blends is, for all practical purposes, independent of the morphologies of the underlying rubber phases. In other words, the nonlinear elastic response of all isotropic 50/50 bicontinuous rubber blends, made with the same rubber phases, is essentially the same. Put yet another way, the nonlinear elastic response of isotropic 50/50 bicontinuous rubber blends is primarily controlled by the volume fraction and bicontinuity of the underlying rubber phases.

In the next step of this investigation, we plan to study the nonlinear viscoelastic response of the same type of blends, as well as of blends with unequal volume fractions of the two rubber phases, by leveraging recent analytical and numerical advances in homogenization within the setting of finite viscoelastostatics \cite{kumar2016two,ghosh2021nonlinear,SGLP23}. Contrary to the nonlinear elastic response studied here, we expect that the nonlinear viscoelastic response of the blends will be significantly dependent on the morphologies of the underlying rubber phases. If so, having a direct link between the macroscopic properties of the blends and their microstructure should prove invaluable to guide the design of new and improved rubber blends.

Another direction worth pursuing is the deployment of the computational tools that we have developed in this work to construct voxelized and simplicial unit cells to study computationally the homogenization of the mechanical (not just elastic and viscoelastic) and physical properties at large of other materials with spinodal microstructures (not just rubber blends). Having access to both the voxelized and simplicial discretizations of the same periodic microstructures should allow to critically explore  the performance of FFT approaches \cite{MS98,PL09,Suquet18,Gelebart20,Schneider23} vis-\`a-vis the FE method for 3D homogenization problems.

\section*{Acknowledgements}

\noindent Support for this work by the tire manufacturing company Michelin is gratefully acknowledged.

\bibliographystyle{elsarticle-num}
\bibliography{refs}

\end{document}